\pdfoutput=1

\documentclass{article}
\usepackage[sort&compress,comma,numbers]{natbib}
\usepackage{arxiv}
\usepackage[utf8]{inputenc} % allow utf-8 input
\usepackage[T1]{fontenc}    % use 8-bit T1 fonts
\usepackage{amsfonts}       % blackboard math symbols
\usepackage{amsmath} 

\usepackage{amssymb}
\usepackage{amsthm}

\usepackage{float}
\usepackage{nicefrac}       % compact symbols for 1/2, etc.
\usepackage{microtype}      % microtypography
\usepackage{graphicx}

\usepackage[mathlines]{lineno}% Enable numbering of text and display math
\usepackage{url}   
\usepackage[pagebackref=true,breaklinks=true,colorlinks,bookmarks=false]{hyperref} 
\usepackage{pdfpages}
\usepackage{color}
\usepackage{cases}

\def\ie{\textit{i.e.}}

\def\mm{Methods}
\def\si{Supplementary Information}

\title{Self-organization of nonlinearly coupled neural fluctuations into synergistic population codes}

\author{
Hengyuan Ma\textsuperscript{1,$\ast$},
Yang Qi\textsuperscript{1,2,}\thanks{These authors contributed equally.}~~~,
Pulin Gong\textsuperscript{4},
Jie Zhang\textsuperscript{1,2}
Wenlian Lu\textsuperscript{1,2,$\dagger$},
Jianfeng Feng\textsuperscript{1,2,3,}\thanks{Correspondence should be addressed to: \texttt{jffeng@fudan.edu.cn} }\vspace{3mm}
\\
\it{1} Institute of Science and Technology for Brain-inspired Intelligence, Fudan University, Shanghai 200433, China
\\
\it{2} Key Laboratory of Computational Neuroscience and
\\
Brain-Inspired Intelligence (Fudan University), Ministry of Education, China \\
\it{3} Department of Computer Science, University of Warwick, Coventry, CV4 7AL, UK \\
\it{4} School of Physics, University of Sydney, Sydney, NSW 2006, Australia\\
\vspace{2em} 
\url{https://github.com/AwakerMhy/moment_nn/tree/main}
}

\begin{document}
\maketitle

\begin{abstract}
Neural activity in the brain exhibits correlated fluctuations that may strongly influence the properties of neural population coding. 
However, how such correlated neural fluctuations may arise from the intrinsic neural circuit dynamics and subsequently affect the computational properties of neural population activity remains poorly understood. The main difficulty lies in resolving the nonlinear coupling between correlated fluctuations with the overall dynamics of the system. In this study, we investigate the emergence of synergistic neural population codes from the intrinsic dynamics of correlated neural fluctuations in a neural circuit model capturing realistic nonlinear noise coupling of spiking neurons. We show that a rich repertoire of spatial correlation patterns naturally emerges in a bump attractor network and further reveals the dynamical regime under which the interplay between differential and noise correlations leads to synergistic codes. Moreover, we find that negative correlations may induce stable bound states between two bumps, a phenomenon previously unobserved in firing rate models. These noise-induced effects of bump attractors lead to a number of computational advantages including enhanced working memory capacity and efficient spatiotemporal multiplexing and can account for a range of cognitive and behavioral phenomena related to working memory. This study offers a dynamical approach to investigating realistic correlated neural fluctuations and insights to their roles in cortical computations. 
\end{abstract}

\section{Introduction}
The firing activity of cortical neurons in the brain exhibits large fluctuations
over time and across trials that are characterized by rich spatiotemporal
correlation structures as revealed by large-scale simultaneous cortical
recordings~\citep{rosenbaum2017spatial,urai2022large}.
It has been suggested that correlated neural variability may play synergistic or destructive roles in neural probabilistic representations such as neural population coding~\citep{kohn2016correlations, panzeri2022structures} and working memory~\citep{doi:10.1073/pnas.1121274109,burak2012fundamental}.
However, existing theoretical analyses of the stochastic dynamics of neural systems primarily rely on firing rate models with additive gaussian noise or simplified Poisson neuron models without correlations~\citep{renart2007mean,burak2012fundamental,doi:10.1073/pnas.1121274109,koyluoglu2017fundamental,wang2022multiple}. As a result, how correlated neural fluctuations may arise from the intrinsic neural circuit dynamics and subsequently affect the computational properties of neural population remains less well understood~\citep{helias2014correlation,dahmen2016correlated}. 
The main difficulty lies in resolving the nonlinear coupling between correlated fluctuations with the overall dynamics of the system.

It is well known that simple structures in a recurrent circuitry can give rise to many emergent properties to neural population activity including symmetry breaking and spontaneous pattern formation~\citep{folias2005breathers,Wavesbumpspatterns}. One such example is found in bump attractors, a type of self-sustained, spatially localized patterns arising from neural circuit models with short-range excitation and long-range inhibition~\citep{amari1977dynamics,Wavesbumpspatterns}. Due to their translation symmetry, bump attractors can be used to encode and store continuous features such as spatial location and orientation~\citep{wu2016continuous}, 
and have been suggested as a possible neural mechanism for a range of perceptual and cognitive processes including spatial navigation~\citep{zhang1996representation,mcnaughton2006path,moser2014grid}, spatial working memory~\citep{compte2000synaptic,wang2001synaptic,wimmer2014bump,zylberberg2017mechanisms},
sensory evidence accumulation~\citep{esnaola2022flexible},
and other cortical functions ~\citep{ben1995theory,seung2000stability}. 
Modeling studies have shown thatmultiple bump states can undergo complex interactions including repulsion, annihilation,merging, and splitting~\citep{qi2015dynamic,krishnan2018synaptic}. 
It has been proposed that neural fluctuations may influence the coding properties of bump attractors in two ways. At short timescales, noisy spike count directly contributes to the coding error of a bump attractor, whereas at long timescales, noise further induces a random drift of the bump attractor along the attractor manifold, resulting in degradation of working memory over time~\citep{burak2012fundamental}. 
Thus, it is critical to understand how correlated fluctuations may arise intrinsically from recurrent neural circuitry and how they affect the dynamics of bump attractors and subsequently the relevant cognitive functions.

In this study, we investigate how synergistic neural population codes
originate from the intrinsic dynamics of correlated neural fluctuations by using a biologically realistic model known as the moment neural network (MNN)~\citep{feng2006dynamics,lu2010gaussian}, which faithfully captures the nonlinear
noise coupling of spiking neurons up to second-order statistical moments~\citep{10.1038/nature06028}.
As suggested by experimental and theoretical studies~\citep{schneidman2006weak,dahmen2016correlated}, second-order statistical moments can provide a minimalistic yet adequate way for capturing the statistical properties of fluctuating neural activity.We construct a bump attractor network based on this model and show that complex spatial correlations naturally emerge from the intrinsic dynamics of recurrent circuits. Analysis of the model reveals the interplay between differential and noise correlations as a key factor determining the computational properties of neural population codes in bump attractors. We also find that negatively correlated neural fluctuations can induce stable bound states between adjacent bumps, a phenomenon previously unobserved in firing rate models. These novel dynamical properties of bump attractors can account for a diversity of experimental observations of cognitive and behavioral responses related to working memory~\citep{van2006eye,theeuwes2009interactions}, and provide a circuit mechanism for efficient neural coding based on spatiotemporal multiplexing~\citep{akam2014oscillatory,caruso2018single} as well as enhanced working memory capacity.

This study offers new insights into how nonlinear noise coupling shapes correlated neural fluctuations into synergistic neural population codes and opens up a dynamical approach to investigating correlated neural fluctuations and their computational properties in neural systems.

\begin{figure}[h]
   \centerline{\includegraphics[width=0.7\textwidth]{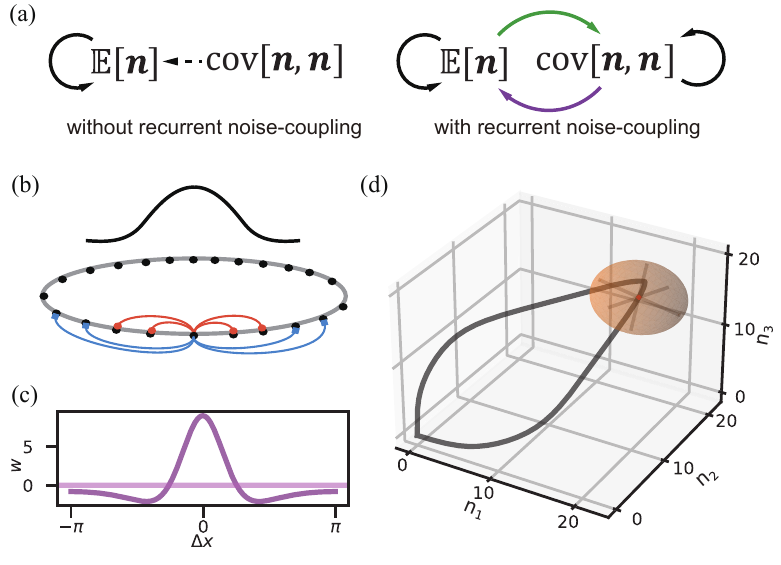}}
	\caption{
	\textbf{Nonlinear noise coupling in a bump attractor network.}
	(a) Comparison of neural circuit models with and without nonlinear noise coupling. In models without nonlinear noise coupling (left panel), the recurrent dynamics affect the spike count mean (solid arrow) with spike count covariance injected additively to the system (dashed arrow). In models with nonlinear noise coupling (right panel), both spike count mean and spike count covariance are coupled recurrently with each other through nonlinear interactions (purple and green arrows). Here, $\mathbf{n}$ represents spike count. 
	(b) A ring network with short-range excitation (red curves) and long-range inhibition (blue curves) supporting bump activity pattern.
	(c) Synaptic strengths $w$ as a function of the distance $\Delta x$ between two neurons.
	(d) The mean firing rate $\mu_i$ of a bump attractor forms a ring manifold (black line) parameterized by its center ofmass, additionally equipped with a firing covariability $C_{ij}$(orange ellipsoid; scaled for visibility) at each point on the attractor manifold. The plot shows the attractor manifold in the space spanned by the activity (spike count) of three neurons.}\label{fig:demo}
\end{figure}

\section{A dynamical model for correlated neural variability}

A major challenge faced by firing rate models is difficulties in capturing realistic neural fluctuations that are nonlinearly coupled to the recurrent circuit dynamics. To illustrate this point, consider the schematic diagram shown in Figure.~\ref{fig:demo}(a). In simplified neural circuit models, neural fluctuation is often modeled as either an external additive input or independent Poisson variability. In a more realistic setting, correlated neural fluctuation may arise intrinsically through the recurrent neural circuit dynamics, as represented by the strong nonlinear coupling between the spike count mean and spike count covariance~\citep{lu2010gaussian,helias2014correlation,dahmen2016correlated}. 
Therefore, to understand the coding and computational properties of bump attractors, it is imperative to develop a model that can accurately describe the statistics and nonlinear coupling of fluctuating neural activity. To overcome this challenge, we employ a model known as the moment neural network (MNN), which faithfully captures the nonlinear coupling of correlated fluctuations in a recurrent population of spiking neurons~\citep{lu2010gaussian}. The model is governed by the following closed system of equations describing the dynamics of the second-order statistical moments of neural activity,
\begin{align}
    &\tau \frac{\partial \mu_i}{\partial t}  = -\mu_i+\phi_{\mu}(\bar{\mu}_i,\bar{\sigma}^2_i),\label{eq:mu_dynamics} \\ 
    &\tau \frac{\partial C_{ij}}{\partial t}= -C_{ij}+\begin{cases}
 \phi_{\sigma}(\bar{\mu}_i,\bar{\sigma}^2_i),&\text{for }i=j,\\
  \psi(\bar{\mu}_i,\bar{\sigma}^2_i)\psi(\bar{\mu}_j,\bar{\sigma}^2_j)\bar{C}_{ij},&\text{for }i\neq j,
\end{cases}\label{eq:C_dynamics}
\end{align}
where the mean firing rate $\mu_i$ and firing covariability $C_{ij}$ represent the mean and covariance of the spike count per unit time, respectively, and $\tau$ is the membrane time constant. 
The functions $\phi_{\mu}$, $\phi_{\sigma}$ and $\psi$ are pointwise activation functions describing the relationship between the input current statistics and the output spike train statistics. For the leaky integrate-and fire (LIF) spiking neuron model, the functional form of the activation functions are given by equations~\eqref{eq:mu} and ~\eqref{eq:sigma} in \mm.
Unlike previous models with heuristic activation functions~\citep{amari1977dynamics,Wavesbumpspatterns,coombes2010neural,wu2016continuous}, these activations are derived through a mathematical technique known as the diffusion approximation~\citep{amit1991quantitative,amit1997dynamics}which faithfully captures the nonlinear coupling of mean firing rate and firing variability across populations of neurons. The quantities $\bar{\mu}_i$ and $\bar{C}_{ij}$ correspond to the mean and the
covariance of the total synaptic current, respectively, and are defined as
\begin{align}
    &\bar{\mu}_i = \sum_{k} w_{ik}\mu_k  + \mu_{\mathrm{ext}}\label{eq:mu_bar}\\
    &\bar{C}_{ij} = \sum_{k,l}  w_{ik}C_{kl} w_{jl}+\sigma_{\mathrm{ext}}^2\delta_{ij},\label{eq:C_bar}
\end{align}
where $w_{ij}$ is the synaptic weight from the $i$th neuron to the $j$th neuron, $\bar{\sigma}^2_i=\bar{C}_{ii}$ denotes the current variability, and $\mu_{\mathrm{ext}}$ and $\sigma^2_{\mathrm{ext}}$ are the mean and variance of a spatially uniform external input, respectively.
We fix $\sigma^2_{\mathrm{ext}}$ throughout this study and systematically vary $\mu_{\mathrm{ext}}$. See \mm~for details of the parameter settings. We note a couple of important features of this model. First, unlike models with linearly coupled covariance as considered in~\citep{buice2010systematic,touboul2011finite}, equations~\eqref{eq:mu_dynamics} and \eqref{eq:C_bar} capture the nonlinear interactions between mean firing rate and firing covariability derived from realistic spiking neuron models. Second, any covariance structure that emerges from the model is due to intrinsic dynamics of the recurrent circuit, not determined by the external input, which is uncorrelated.

To implement the bump attractor, each neuron $i$ is assigned to spatial
coordinates $x_i\in[-\pi,\pi)$ over a feature space with a periodic boundary condition and is connected to each other via synaptic connections with short-range
excitation and long-range inhibition according to equation~\eqref{eq:weight} in \mm. A schematic diagram of this neural circuit structure is shown in Figure~\ref{fig:demo}b, and the synaptic connections are shown in Figure~\ref{fig:demo}c.

\section{Correlated variability in bump attractors}
In firing rate-based models without correlation, distance-dependent synaptic coupling such as that shown in Figure~\ref{fig:demo}b can give rise to stable activity state known as bump attractors~\citep{amari1977dynamics}. Due to the translation
invariance of the synaptic coupling, the mean firing rate $\mu_i$of a bump attractor forms a ring manifold parameterized by its center of mass. Experimentally, bump or ring attractors have been found in various biological neural systems such as the fruit fly ellipsoid body~\citep{kim2017ring}, primate V1~\citep{rosenbaum2017spatial}, and primate prefrontal cortex~\citep{wimmer2014bump}. 
However, the firing rate–based model is inadequate, since it omits the
fact that neurons in the ring attractors are correlated (e.g., Fig.7 in ~\citep{wimmer2014bump}). The MNN overcomes this shortcoming by integrating the ring manifold with the intrinsic covariance structure of neural activity that is nonlinearly coupled with the mean firing rate. As a result, the topology of the attractor manifold in our model is no longer represented by points on a ring embedded in the Euclidean space $\mathbb{R}^n$ spanned by the mean firing rates $\mu$, but also extends to the positive semidefinite cone $S^n_{+} (\mathbb{R})$ spanned by the firing covariability $C$. This topological structure can be visualized as a family of ellipsoids with their centers located on the ring manifold and with their sizes and orientations representing the covariance structure of neural activity (Figure~\ref{fig:demo}d).

\begin{figure}[t]
    \centering
    \includegraphics[width=0.7\textwidth]{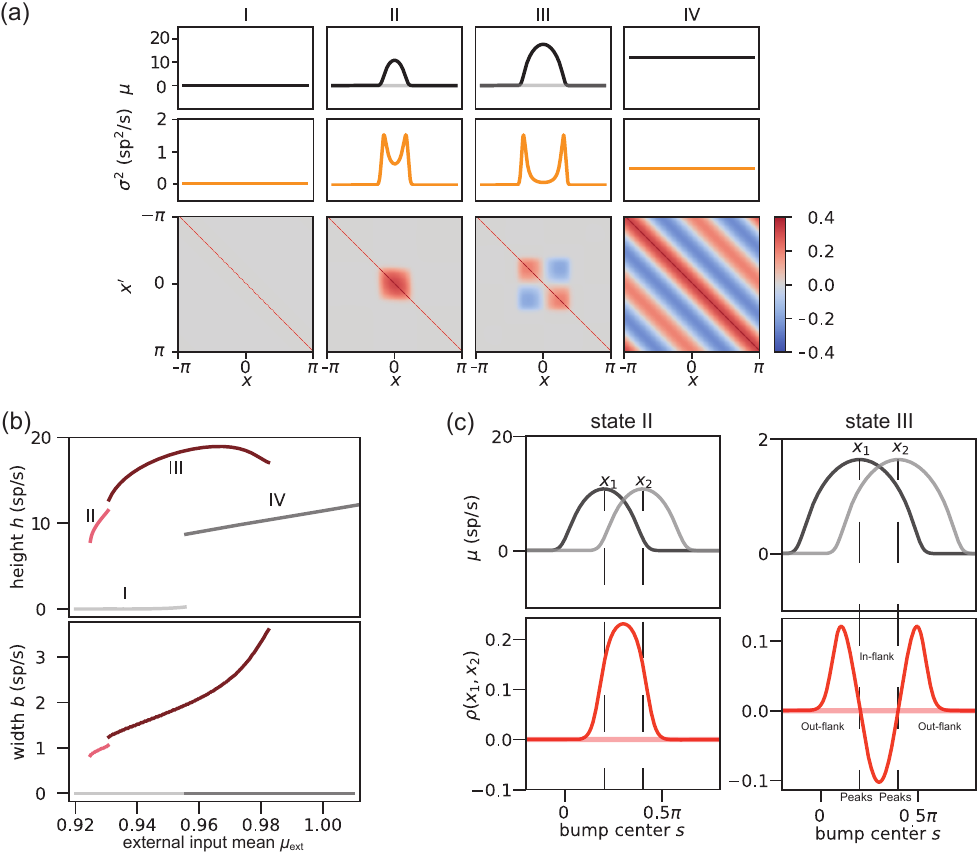}
    \caption{\textbf{Correlated variability emerging from a bump attractor network.}
    (a) Mean firing rate $\mu(x)$, firing variability $\sigma^2(x)$ and the correlation coefficients $\rho(x,{x}')$ of neural activity states for varying external input mean $\mu_{\mathrm{ext}}$, including spatially uniform resting state (state I, $\mu_{\mathrm{ext}}=0.920$), bump state with positive correlations (state II, $\mu_{\mathrm{ext}}=0.929$), bump state with both positive and negative correlations (state III, $\mu_{\mathrm{ext}}=0.948$), and globally active state with spatially periodic correlations (state IV, $\mu_{\mathrm{ext}}=0.986$).
    (b) The height $h$ (top panel) and width $b$ (lower panel) for varying external input mean $\mu_{\mathrm{ext}}$, revealing distinct phase transitions between different activity states shown in panel a.
    (c) The correlation coefficient $\rho$ between two neurons with preferred features $x_1$ and $x_2$ as the bump location $s$ varies in state II and state III.}
    \label{fig:simple_bump_demo}
\end{figure}

In the following, we reveal the emergence of intrinsic spatial correlations in the bump attractor network through numerical simulations of the system defined by equations~\eqref{eq:mu_dynamics}-\eqref{eq:C_bar}. 
Similar to conventional firing rate–based models~\citep{amari1977dynamics,Wavesbumpspatterns}, when the external input mean $\mu_{\mathrm{ext}}$ is weak, the system has only one stable solution, the resting state (see state I in Figure~\ref{fig:simple_bump_demo}a).
As $\mu_{\mathrm{ext}}$ increases, stable bump states emerge (see State II in Figure~\ref{fig:simple_bump_demo}a). Consistent with previous findings~\citep{amari1977dynamics,Wavesbumpspatterns}, the mean firing rate $\mu_i=\mu(x_i\vert s)$ of the bump state exhibits a unimodal spatial profile peaked at its center $s$. In contrast, the firing variability $\sigma^2_{i}=\sigma^2(x_i\vert s)$ is bimodal, with two peaks near the edges of the bump and a local minimum at the center. This is in marked contrast to the commonly assumed independent Poisson activity, in which the firing variability is proportional to the mean firing rate or uncorrelated gaussian noise with constant magnitude~\citep{ma2010signal,burak2012fundamental}.
We find that neural activity is positively correlated within the bump and uncorrelated outside it. Note that the stable bump state coexists with the stable resting state, allowing the system to switch between them when subjected to appropriate external stimuli.

Furthermore, unlike firing rate–based models, additional phase transitions occur in the correlation structure of the bump attractor. As $\mu_{\mathrm{ext}}$further increases, negative correlations emerge between neurons on the opposite sides of the bump, while the shape of the mean firing rate remains qualitatively the same (see state III in Figure~\ref{fig:simple_bump_demo}a). Note that both the mean and variance of the external input are uniform, suggesting that the spatial correlation structure is an emergent property of the system. Similar to firing rate–based models, further increasing $\mu_{\mathrm{ext}}$eventually leads to the saturation of mean firing rate into a spatially uniform activity. However, we find that the same case is not true for spatial correlations, which transform into a periodic wave pattern (see state IV in Figure~\ref{fig:simple_bump_demo}a). Such a distance-dependent
correlation structure may potentially arise from synchronized firing with the same and the opposite phases.

To quantify the transition between these different activity patterns, we calculate the height $h$ and the full width at half maximum $b$ of the bump solution for varying the external input mean $\mu_{\mathrm{ext}}$. As shown in Figure~\ref{fig:simple_bump_demo}b, the phase diagrams constructed from these measurements reveal four distinct regimes occupied by the neural activity patterns I-IV. We find that both $h$ and $b$ vary smoothly as a function of $\mu_{\mathrm{ext}}$ within each regime and that they exhibit discontinuous jumps at phase transitions. The discontinuous jumps are qualitatively consistent with saddle-node bifurcations found in firing rate bump attractor models~\citep{amari1977dynamics,Wavesbumpspatterns}.
It is worth noting that both the mean firing rate and the correlation coefficients exhibited by our model are within a biologically plausible range, consistent with experimental observations~\citep{wimmer2014bump}.
 
To further illustrate the internal correlation structure of the bump attractor, we focus on two neurons and inspect how their mean firing rate and correlation depend on the bump center $s$. Specifically, given two neurons located at $x_1=0.2\pi$ and $x_2=0.4\pi$, their mean firing rates $\mu$ follow bell-shaped tuning functions peaked at $x_1$ and $x_2$ respectively. This holds for both states II and III, as shown in the top panels in Figure~\ref{fig:simple_bump_demo}(c). For each state, a distinct pattern is found in the correlation coefficient between these two neurons. For state II, the correlation coefficient $\rho(x_1,x_2)$ reaches the maximum when s is located in the middle between $x_1$ and $x_2$, and monotonically decays to zero when s move away from the middle point in both directions.
For State III, when $x_1<s<x_2$ (in-flank cases), we have $\rho(x_1,x_2)<0$; when $s=x_1$ or $s=x_2$ (peak cases), we have $\rho(x_1,x_2)=0$; and when $s<x_1$ or $s>x_2$ (out-flank cases), we have $\rho(x_1,x_2)>0$.
We find that the correlation pattern predicted by state III of our model is consistent with experimentally observed pairwise correlation structures in the primate prefrontal cortex~\citep{wimmer2014bump}. This study analyzed neuronal recordings in monkeys during a visuospatial delayed response task, in which they were required to memorize the spatial locations of transient stimuli and found that the tuning curves and pairwise correlation structures in the prefrontal cortex corroborated the bump attractor hypothesis, that is, the neural population held a bump activity that encoded the cue location $s\in [-\pi,\pi)$ by the bump center during the delay period. As shown in Figure 7f of their paper, in the out-flank, in-flank and peak cases, the sign of the correlations of two neurons with two preferred stimuli $x_1$ and $x_2$ are the same as our prediction in State III of Figure~\ref{fig:simple_bump_demo}c.
Additionally, our model predicts the existence of a
pairwise correlation structure as in state II of Figure~\ref{fig:simple_bump_demo}c, which has not been experimentally observed before.

Importantly, in both~\citep{wimmer2014bump} and our model,the correlations of each neuronal pair are dependent on the location of the bump center $s$, and this is more general than previous models, where the correlations are assumed to be translation invariant~\citep{shamir2004nonlinear,averbeck2006neural,ecker2011effect}.
Although both the rate-based model~\citep{wimmer2014bump}and theMNN display a bell-shaped firing rate profile
during the delay period of working memory (compare Figure 8a in their paper with State III in Figure~\ref{fig:simple_bump_demo}), the MNN offers a number of unique advantages.
First, in the rate-based model, the spiking activity of each neuron
is modeled as an inhomogeneous Poisson process whose instantaneous firing rate is determined by the dynamics of the rate-based model. In contrast, the MNN directly captures the dynamics of neural spiking activity in terms of the nonlinear coupling between mean firing rate and firing covariability, without imposing such simplifying abstractions. Moreover, in the rate-based model, the diffusion of bump state along the attractor manifold due to external noise is required to explain several experimental observations such as negative noise correlation and stimulus-dependent Fano factor. However, by considering the intrinsic covariance structure in the MNN, many of these experimental observations can be explained without invoking the diffusion of bump state. See the supplementary information (\si) for detailed comparisons between the stochastic rate-based model and the MNN. These analyses demonstrate that by considering the nonlinear noise coupling in recurrent neural circuits, the MNN can account for the correlation structures of neural fluctuations in experimental data and lay the theoretical foundation for exploring the role of correlated neural fluctuations in brain functions such as neural coding and working memory.

\section{Noise covariance influences coding accuracy}
\begin{figure}[h]
	\centering
   \centerline{\includegraphics[width=0.5\textwidth]{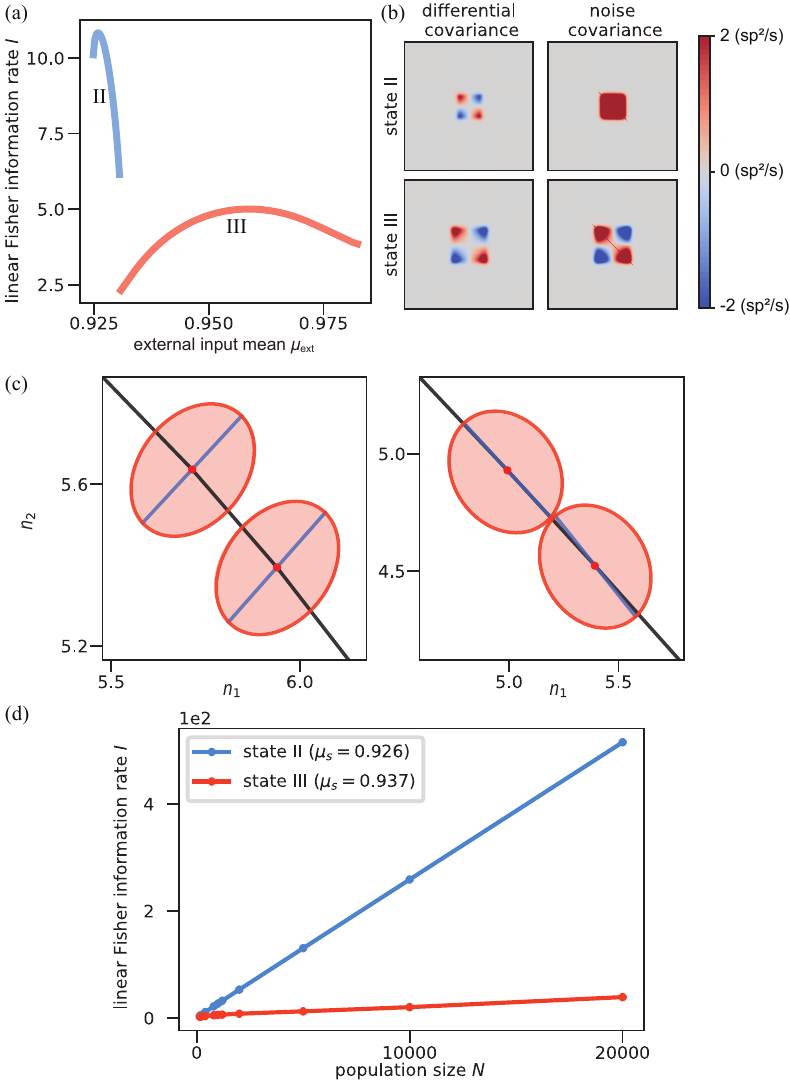}}
    \caption{\textbf{Noise and differential covariance together influence coding accuracy.} (a)
    Linear Fisher information rate $I$ under varying external input mean $\mu_{\mathrm{ext}}$. (b) Differential covariance and noise covariance of state II ($\mu_{\mathrm{ext}}=0.925$) and state III ($\mu_{\mathrm{ext}}=0.931$).
    (c) Projecting the system state onto the space spanned by the activity of two neurons whose preferred $s$ differs by $0.2\pi$.
    The black line
    shows the ring manifold represented by the the mean firing rates of two neurons.
    Orange ellipses represent the firing covariability when the bump center is
    located at different values represented by the red dots. Both state II (left panel)
    and state III (right panel) show similar structures in mean firing rate, but not
    the firing covariability. The major axis (blue lines) of the firing covariability in
    state II is roughly orthogonal to the ring manifold, whereas that in state III is
    roughly parallel to the ring manifold.
    (d) Linear Fisher information rate $I$ grows linearly with the neural population size $N$.
    }
    \label{fig:coding}
\end{figure}

Due to the translation symmetry of bump attractor, its center s can be used to encode continuous variables. We analyze how the coding accuracy of bump attractor is influenced by correlated activity state. The coding accuracy can be quantified using the linear Fisher information$I_{\mathrm{linear}}(\Delta t)= \mathbf{u}(s)^TC(s)^{-1}\mathbf{u}(s)\Delta t$, 
where $\mathbf{u}_i(s) = \frac{d \mu_i(s)}{d s}$corresponds to the tangent vector along the attractor manifold and $\Delta t$ is the spike count time window.
The linear Fisher information $I_{\mathrm{linear}}(\Delta t)$ is approximately independent of $s$ for large population size due to translation invariance. Linear Fisher information provides an upper bound on the certainty of the bump’s location as can be determined by an optimal linear decoder from random spike count~\citep{graf2011decoding,berens2012fast,fetsch2012neural}.
Figure~\ref{fig:coding}a shows the linear Fisher information rate ($I:=\frac{I_{\text{linear}}(\Delta t)}{\Delta t}$) of a bump attractor for encoding $s$ under varying the external input mean $\mu_{\mathrm{ext}}$. 
We observe that the phase transition from state II to state III caused by increasing $\mu_{\mathrm{ext}}$ coincides with an abrupt decrease in $I$.
Since cognitive performance at the behavioral level is limited by the coding accuracy in upstream neural populations~\citep{li2021joint}, these two dynamical regimes of bump activity state may correspond to different levels of cognitive performance. In particular, our model predicts that variations in global circuit parameters such as task-independent external input mean $\mu_{\mathrm{ext}}$are sufficient to induce shifts in cognitive state, which may manifest as a sudden deterioration of task performance.

To understand what causes the drastic difference in the coding accuracy between bump states II and III, we turn to analyze the contribution of correlated neural variability to the linear Fisher information. We find that the linear Fisher information is jointly influenced by two types of firing covariability, differential covariance and noise covariance~\citep{averbeck2006neural}. Differential covariance $\mathbf{u}(s)\mathbf{u}(s)^T$ measures the covariance due to changes in the mean firing rate $\mu$ under infinitesimal shifts in $s$~\citep{kohn2016correlations}, whereas noise covariance $C(s)$ in our model measures the spike count covariance per unit time between two neurons for fixed $s$. Figure~\ref{fig:coding}b compares the spatial structures of differential covariance and noise covariance in states II and III. Both states exhibit similarly shaped differential covariance with positively correlated neurons at the same side of the bump and negatively correlated neurons at the different sides of the bump, consistent with previous firing rate–based models of bump attractor~\citep{wu2016continuous}.
However, our model reveals distinct shapes in the noise covariance
that could not be captured by previous models. For state II, the noise covariance is nonnegative, whereas for state III, the noise covariance contains both positive and negative values. It turns out that unlike state II, both noise and differential covariances have the same sign in state III, which could cause the sudden decrease in linear Fisher information. This result is consistent with previous theoretical analysis of neural population coding~\citep{moreno2014information}, which attributes the information decrease to the overlap between these two kinds of covariances.

The interplay between the differential covariance and noise covariance can be further understood geometrically by considering two-dimensional projections of the bump attractor state onto the space spanned by the spike count of two neurons. As shown in Figure~\ref{fig:coding}c, the black lines represent the mean firing rate of the bump attractor in the space spanned by the activity of two neurons whose preferred $s$ differ by $0.2\pi$.
Orange ellipses represent the firing covariability (scaled for visibility) for when the mean firing rates take specific values on the ring manifold as marked by the red dots. In this figure, the differential covariance corresponds to the tangent space of the attractor manifold represented by the mean firing rate, whereas noise covariance corresponds to the orange ellipses. Remarkably, the major axis (blue line) of the noise covariance is roughly orthogonal to the attractor manifold in state II but is roughly parallel to the attractor manifold in state III. As a result, the coding error, which is dominated by the projection of the noise covariance onto the attractor manifold, is smaller in state II than in state III.

Previous theoretical studies on neural population coding suggest that Fisher information may saturate with increasing neural population size due to information-limiting correlations~\citep{averbeck2006neural,moreno2014information,panzeri2022structures}. It is shown analytically that in a neural population with homogeneous tuning function and distance-dependent noise correlation $\rho(x_1,x_2)=\rho(x_1-x_2)$, Fisher information grows linearly or superlinearly with population size when the noise correlation is zero or negative, respectively, but saturates to a finite value when the noise correlation is positive~\citep{sompolinsky2001population}. However, this saturation may disappear when the neural population has heterogeneous tuning~\citep{ecker2011effect}. It is pointed out that such information-limiting correlation can be largely attributed to the overlap between noise correlation C and differential correlation $\mathbf{u}\mathbf{u}^T$ ~\citep{moreno2014information}.

We run simulations of the neural circuit model with varying numbers of
neurons to investigate the effect of population size on linear Fisher information rate $I$ for states II and III, respectively (see Figure~\ref{fig:coding}d). 
We find that in both cases, linear Fisher information grows approximately linearly with the population size, though this growth is severely reduced in the case of state III. This outcome can be explained by observing that the noise covariance in state III significantly overlaps with differential covariance but not in state II (see Figure~\ref{fig:coding}b). While information is reduced in state III, it does not saturate to a finite limit as the noise covariance is similar in sign but not completely parallel to the differential covariance. This is confirmed by calculating the ratio between the information and the population size, which converges to a nonvanishing value in the limit of large population size (\si, including Figure~S1),and is further elucidated by theoretical analysis using simplified covariance structures (see equations S1 and S30 in \si).

These results are thus consistent with the theory that differential correlation serves as an information-limiting correlation~\citep{moreno2014information}. Furthermore, our model captures the emergence of mean firing rate $\mu(x)$ and noise covariance $C(x,{x}')$ through the intrinsic dynamics of the recurrent circuit, thereby offering an explanation about the dynamical origin of information-limiting correlations. Importantly, our model demonstrates that such information-limiting correlation can be switched on and off by changing a global parameter (the external input mean $\mu_{\mathrm{ext}}$ in this case), which determines the dynamical regime of the bump attractor.

Previously, it has also been found that the information content in a bump attractor is more concentrated around the edges of the bump state with independent Poisson activity~\citep{seung1993simple}.
We also investigate the contribution of linear Fisher information from each neuron as shown in \si, Figure S2. We find that in state II, the neurons located at the boundaries of the bump are more important for coding than other neurons, as consistent with previous models~\citep{seung1993simple}.
Unexpectedly, in state III, the neurons near the center of the bump provide negative contributions to the linear Fisher information, suggesting that some neurons may be harmful for coding. Since the correlated fluctuation is coupled with the neural dynamics in our model, our findings suggest a way for the neural population to directly regulate its coding efficiency through manipulating its intrinsic correlation structure.

The analysis above primarily focuses on the coding property of stationary bump activity. In the context of working memory, this can be thought as representing the neural activity state during the delay epoch of working memory. In addition to such time-independent mnemonic coding subspace, \cite{murray2017stable} have also identified the presence of a dynamic coding subspace by analyzing neural spike data of the primate prefrontal cortex during working memory tasks.
Following~\cite{murray2017stable}, we investigate the dynamic coding in our MNN by analyzing the temporal dynamics of the bump state during and after cue presentation for each of states II and III (see Figures S5-S6 in \si).
We observe that the population correlation
(equation S1 of~\citep{murray2017stable}) between the population activity states and a reference population activity at the cue epoch reaches its peak during the cue epoch and gradually decays during the delay epoch (see Figure S5c, \si). This observation suggests the existence of a dynamic subspace that undergoes decay during the delay epoch. Furthermore, we find that the linear Fisher information of the dynamic coding in state II (see Figure~\ref{fig:simple_bump_demo})is significantly higher than that of the mnemonic coding, and the linear Fisher information of both types of coding becomes similar during the delay epoch (see Figure S5d, \si).
This is consistent with that study, which demonstrated that the decoding accuracy of dynamic coding is substantially higher than that of mnemonic coding during the cue epoch, but both accuracies become similar during the delay epoch. Interestingly, we find the opposite result in state III (see Figure~\ref{fig:simple_bump_demo}), where the linear Fisher information of mnemonic coding is significantly higher than that of the dynamic coding during the cue period. This suggests that the brain might adopt a correlation structure in state II during the cue epoch for better coding efficiency. (See the \si~for details.)

\begin{figure}[h]
\centering
\includegraphics[width=0.7\textwidth]{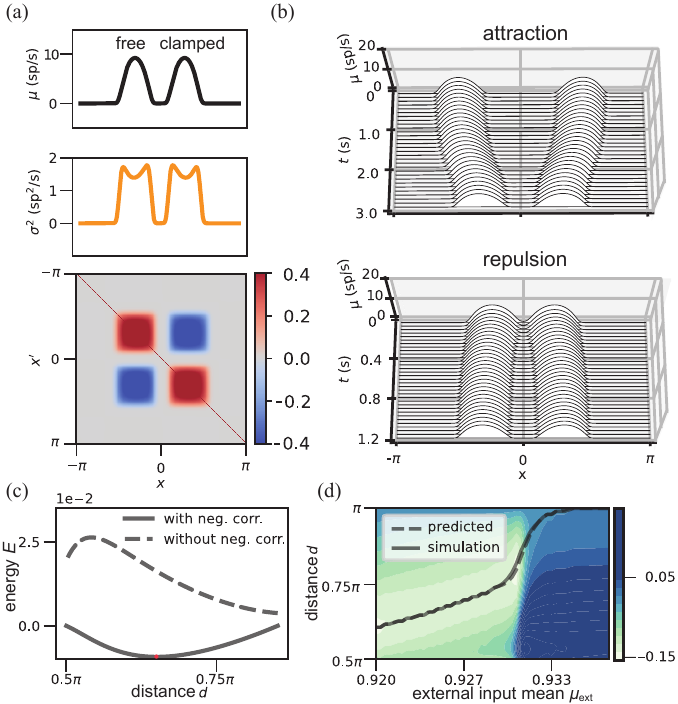}
    \caption{\textbf{Interactions of bump attractors mediated by negatively correlated fluctuations. }
    (a) Mean firing rate $\mu$, firing variability $\sigma^2$ and correlation coefficient $\rho$ of two bumps in stable bound state.
    (b) Two bumps separated by a distance closer to or farther from the stable equilibrium at $d^{\ast}\approx 0.63\pi$ experience repulsion or attraction, respectively.
    (c) The effective interaction energy $E$ as a function of interbump distance $d$ (solid line) exhibits a local minimum at $d\approx0.65\pi$ (dot), which disappears when the negative correlation is removed (dashed line). We set $\mu_{\mathrm{ext}}=0.922$ in panels a-c.
    (d) Stable equilibrium distance $d^\ast$ between two bumps for varying external input mean $\mu_{\mathrm{ext}}$ (black lines), overlaid on the effective interaction energy $E$ (heat map).
    }
    \label{fig:2_bump_compre}
\end{figure}

\section{Stable bound states of bump attractors}
In the previous section, we revealed the influence of correlated neural variability on neural coding by a single bump attractor state. In this section, we further investigate its impact on the dynamical interaction of multiple bump attractor states. In bump attractor networks, multiple bump states elicited by different stimuli may coexist in the same neural circuit and interact through distance-dependent synaptic connections. Conventional firing rate–based models predict that such interactions typically lead to either repulsion due to long-range synaptic inhibition or merging due to short-range excitation~\citep{krishnan2018synaptic,wojtak2021dynamic}.

In our model, however, we find that bump attractors may become
dynamically coupled through negatively correlated fluctuations, forming
stable bound states previously unobserved in firing rate–based models.
Figure~\ref{fig:2_bump_compre}a shows two bumps (see State II in Figure~\ref{fig:simple_bump_demo}a) forming such a
stable bound state. Both the mean firing rate $\mu$ and firing variability $\sigma^2$ of each of them largely retain the similar shapes as that of a single isolated bump. However, the system now also exhibits negative correlations between two bumps, in addition to the positive correlations within each bump. Remarkably, the interbump distance $d$ of such a two-bump state forms a stable equilibrium, that is, if the system is initialized with the interbump distance $d$ smaller or larger than the stable equilibrium, the bumps will repulse or attract each other to eventually reach the stable distance, as shown in Figure~\ref{fig:2_bump_compre}b. 
Similar phenomena of stable bound states have been found in other noise-coupled nonlinear systems such as interacting solitons in mode-locked lasers~\citep{grelu2012dissipative,weill2016noise} and electron-electron attraction due to lattice vibration in superconductors~\citep{bardeen1957microscopic}.

To quantify the interaction between two bumps, we employ a projection method to calculate the effective interaction energy. In this method, we first clamp one of the two bumps in Figure~\ref{fig:2_bump_compre}b and consider its influence on the other bump (the free-moving bump) as if it acts like an external input. We next assume that the shape of the free-moving bump does not vary significantly as it is shifted around the stable equilibrium $s^{\mathrm{f}}$ by $z$. 
Under these simplifications, we can project the entire system state on to the neutrally stable left eigenspace along the attractor manifold represented by the mean firing rate of the free-moving bump to arrive at an approximate equation of motion for the bump’s spatial deviation $z$~\citep{burak2012fundamental}. 
Due to the complexity of our model, explicit analytical solution is prohibitive, and we instead resort to numerical analysis as follows. We approximate the mean firing rate and firing covariability of the free bump as $\mu^{\mathrm{f}}_i(z)={\mu}^{\mathrm{f}}(x_i\vert s^{\mathrm{f}}+z)$ and $C^{\mathrm{f}}_{ij}(z)={C}^{\mathrm{ff}}(x_i,x_j\vert s^{\mathrm{f}}+z)$, respectively, where ${\mu}^{\mathrm{f}}(x \vert s^{\mathrm{f}})$ and ${C}^{\mathrm{ff}}(x,{x}' \vert s^{\mathrm{f}})$ represent the shape of the free bump's mean and covariance at the stable equilibrium $s^{\mathrm{f}}$.  
Following~\cite{touboul2011finite}, we denote 
$\langle f,g\rangle=\tfrac{2\pi}{N}\sum_{i}f(x_i)g(x_i)$ as the inner product, and calculate the left eigenvector of the system for the free bump near the equilibrium as $v = a\frac{d\bar{\mu}^{\mathrm{f}}}{dz} $, where $\bar{\mu}^{\mathrm{f}}$ is the synaptic current received by the free bump (equation S37), and $a=1/\langle \frac{d\bar{\mu}^{\mathrm{f}}}{dz},\frac{d\mu^{\mathrm{f}}}{dz} \rangle$ is a normalization coefficient.
Projecting both sides of equation~\eqref{eq:mu_dynamics} for the full system to the left-eigenspace spanned by $v$, we obtain 
\begin{align}
    \frac{dz}{dt} = \left \langle  v ,-\mu(z) +\phi\big(\bar{\mu}(z),\bar{C}(z)\big) \right \rangle := b(d),
\end{align}
where $\mu(z)$ and $C(z)$ denote the system state after shifting the free bump by $z$, $d=s^{\mathrm{c}}-s^{\mathrm{f}}-z$ is the interbump distance, and $s^{\mathrm{c}}$is the peak location of the clamped bump. Based on this equation of motion, given the distance $d$ between two bumps, we can define the corresponding effective interaction energy as $E(d) = -\int_{\pi}^d b(x)dx$. (See equations S32 to S46 for details of the derivations of$b(d)$.) As shown in Figure~\ref{fig:2_bump_compre}c, the stable bound state of bump attractors corresponds to the local minimum of the effective interaction energy with an interbump distance of $d^\ast=s^{\mathrm{c}}-s^{\mathrm{f}}=0.65\pi$.
Interestingly, when the negative correlation between two bumps is clamped to zero, the bound state loses its stability as indicated by the disappearance of the local minimum in the effective interaction energy (see Figure~\ref{fig:2_bump_compre}c), resulting in repulsive interaction. 
We find that as we increase the strength of the external
input mean $\mu_{\mathrm{ext}}$, the stable distance $d^{\ast}$ between two bump attactors increases, as shown by the black curve in Figure~\ref{fig:2_bump_compre}d.
Interestingly, the stable distance $d^{\ast}$ exhibits a steep increase near $\mu_{\mathrm{ext}}=0.93$, which coincides with the transition point between states II and III of a single bump in Figure~\ref{fig:simple_bump_demo}.
However, the internal covariance structures of these two bumps do not change until $\mu_{\mathrm{ext}}$is much higher, presumably due to the interaction between the bumps (see Figure S3).
The stable distance curve in Figure~\ref{fig:2_bump_compre}d is laced over the effective interaction energy, whose local minimum for each fixed $\mu_{\mathrm{ext}}$ coincides with $d^{\ast}$.
These results indicate that negative interbump correlation is responsible for generating the stable-bound states of bump attractors in our model, preventing strong repulsion between two bumps as typically observed in firing rate-based models.

To gain further theoretical understanding of the origin of negative interbump correlations, we consider the overall interactions between the neural populations of these two bumps, which can be roughly characterized by two neural masses with self-excitation and mutual inhibition. Denote the mean firing rate and firing covariability of these two neural masses at the steady state as
\begin{align}
    \mu = \begin{pmatrix}
            \mu_1 \\
            \mu_2
            \end{pmatrix},
    \quad C = \begin{pmatrix}
            c_{11}&c_{12}  \\
            c_{21}&c_{22}  
            \end{pmatrix},
\end{align}
and the average synaptic weights between the two populations as $w = \bigl( \begin{smallmatrix}w_{11}&w_{12}\\ w_{21}&w_{22}\end{smallmatrix}\bigr)$, such that
$w_{11},~w_{22}>0$ and $w_{12},~w_{21}<0$, as required by the self-excitation and the mutual inhibition. Under mild conditions over the neural activation, which are satisfied in our model, the activity of these two populations of neurons becomes negatively correlated: $c_{12}<0$ (see equations S47 and S55 for proof). This implies that the negative correlation is a direct consequence of the inhibition between two neural populations. This analysis leads to the surprising conclusion that mutual inhibition between bumps increases working memory capacity by reducing the minimal distance between adjacent bumps, contrary to conventional wisdom suggesting that inhibitions should limit working memory capacity through repulsive interactions
between bumps~\citep{edin2009mechanism,krishnan2018synaptic}. We turn to this point in the next section.

\section{Intrinsic negative correlations regulate dynamical interactions between memorized items}
 
Previous studies using a firing rate–based model found that the interactions between two bumps are attractive when they are close, and they become repulsive past an unstable equilibrium~\citep{wojtak2021dynamic}, similar to the energy landscape indicated by the dashed line in Figure~\ref{fig:2_bump_compre}c.
However, such attractive or repulsive interactions will either cause the two bumps to merge, preventing the system from discriminating two distinct items~\citep{almeida2015neural},or to drift away from each other, thereby limiting the coding resolution. First, our model demonstrates that through negative correlations, two bumps can maintain a closer distance than the case without negative correlations. Moreover, when two bumps are too close, the repulsive interaction will restore their distance to the stable equilibrium. As a result, our findings provide a mechanism enabling discrimination of similar items encoded by bump states~\citep{lin2009influence}. 
Second, our results suggest that nonlinearly coupled, correlated noise can have unexpected effects of stabilizing the spatial location of bump states, in contrast to previous findings that bump states should drift away along the attractor manifold under the influence of additive uncorrelated noise~\citep{burak2012fundamental}. In particular, when the position of one bump is maintained through external mechanisms such as external stimulus or top-down attention, another adjacent bump state can remain stabilized through the attraction toward the equilibrium distance. Our model thus provides a potential way to improve the coding accuracy of a bump attractor against this drifting effect.

The preceding analysis of bump interactions can be applied to understand a range of experimental observations of cognitive and behavioral processes involved in working memory~\citep{wimmer2014bump,wu2016continuous} and eye movement presumably reflecting higher cognitive states~\citep{van2006eye,theeuwes2009interactions,wojtak2021dynamic}. 
It has been widely observed that the trajectory of saccadic eye movement (typically from an initial fixation to a target location) can deviate either toward~\citep{van1987stimulus} or away from~\citep{theeuwes2005remembering}an element other than the target in visual field. In one study~\citep{van1987stimulus},human participants were instructed to perform two saccades from fixation toward different targets in consecutive trials. When the delay between two trials was short and the distance between two targets was large, the trajectory of the second saccade deviated toward the first target instead of directly going to the second target through a straight line. In another study~\citep{theeuwes2005remembering}, on the contrary, the trajectory of eye movement deviated away from a nearby visual stimulus whose location was required to be memorized. Existing theoretical proposals for explaining these phenomena involve top-down modulation whose ability to control rapid eye movements, however, appears to be limited both temporally and spatially~\citep{van2009limits}, and an exact mechanism is still unclear~\citep{theeuwes2009interactions}.

The stable bound states found in our model provide an alternative neural mechanism for the above phenomena without resorting to top-down modulation. We first establish the connection between our model and the above two experiments. In our bump attractor model, the ring attractor manifold represents the angular direction in visual space, whereas the previously memorized and the current targets of eye movement are encoded by the centers of bump activities. The clamped bump in our model as in Figure~\ref{fig:2_bump_compre}correspond to the direction of the previously shown target in~\citep{van1987stimulus} or that of the stimulus stored in working memory in~\citep{theeuwes2005remembering}.  
In both cases, the clamped bump induces an energy landscape
like that in Figure~\ref{fig:2_bump_compre}(c),which interacts with the free-moving bump encoding the angular direction of the affected eye movement. It turns out that the influence of the energy landscape induced by the clamped bump is consistent with these experiments: the interaction of the two bumps is attractive when their distance is large, which can explain the deviation of eye movement toward the previously shown target in~\citep{van1987stimulus},and is repulsive when their distance is small, which can explain the deviation of the eye movement away from the stimulus location in~\citep{theeuwes2005remembering}.
The nontrivial dependence of bump interaction on their distance as predicted by our model can be tested by future experiments through systematically investigating the dependence of behavioral errors during working memory tasks on the difference between memorized items or between one memorized item and a distractor stimulus. These results demonstrate that negatively correlated neural fluctuations modulate the interaction between bump states, which in turn influences cognitive functions and behavior and highlights the importance of considering correlated neural fluctuations in modeling neural circuit dynamics. In the next section, we analyze how these noise-mediated dynamical effects can facilitate neural representation in working memory.

\section{Intrinsic negative correlations lead to efficient coding strategies for working memory}
Due to their translation invariance and bistability, bump attractors have been proposed as the neural substrate for spatial working memory~\citep{wimmer2014bump,zylberberg2017mechanisms,wu2016continuous,doi:10.1073/pnas.1121274109,edin2009mechanism}. The number of bumps the system can simultaneously maintain is interpreted as working memory capacity. To investigate how working memory capacity is influenced by correlated neural fluctuations, we elicit multiple bump states in our model with bell-shaped
transient inputs.

\begin{figure}[h]
    \centerline{\includegraphics[width=0.7\textwidth]{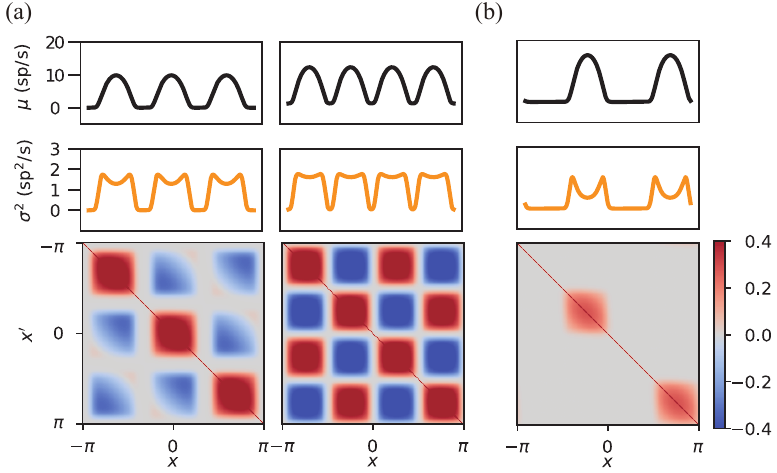}}
    \caption{\textbf{Multiple bump attractors with intricate correlation structures.}
    The mean firing rate, firing variability, and correlation coefficient of the neural circuit with three (left panels) or four (right panels) bumps. In the three-bump case, correlations between different bumps are mainly negative. In the four-bump case, adjacent bumps are negatively correlated, while every other bump is positively correlated. (b) The circuit fails to support more than two bumps when the negative correlation is clamped to zero.}
    \label{fig:multi_bump_demo}
\end{figure}

Neural activity states holding multiple (three and four) bumps (see state II, $\mu_{\mathrm{ext}} =0.935$) are shown in Figure~\ref{fig:multi_bump_demo}a.  The shape of mean firing rate $\mu$ and firing variability $\sigma^2$ of individual bumps are qualitatively consistent with that of an isolated bump (Figure~\ref{fig:simple_bump_demo}), but with slight deformations, lower mean firing rate, and higher firing variability.
When the number of bumps is three, correlations with different bumps are mainly negative, similar to that in a two bumps’ case (see Figure~\ref{fig:2_bump_compre}). When the number of bumps is four, adjacent bumps are negatively correlated (spaced by $\pi/2$), while every other bump is positively correlated (spaced by $\pi$). As we have shown with the two-bump case, negative correlation can reverse the strong repulsion between adjacent bumps, suggesting that negative correlation could be a key factor for enabling the multibump configuration shown above. To test this idea, we clamp the negative correlations in the four-bump case in Figure~\ref{fig:multi_bump_demo}a to zero and find that the activity of some bumps is quenched by the mutual inhibition from adjacent bumps, resulting in the failure for the system to hold more than two bumps, as shown in Figure~\ref{fig:multi_bump_demo}b. This observation is consistent with that the removal of negative correlation turns the interaction between two bumps from a balance of attraction and repulsion into pure repulsion, as described in Figure~\ref{fig:2_bump_compre}.  These results suggest that negative correlations can enable multiple bumps to become more closely packed, thereby enhancing the memory capacity of the system.

Experimental evidence suggests that working memory capacity is limited~\citep{cowan2001magical,cowan2010magical}. There are two theoretical mechanisms through which such limit may occur. The first is a spatial mechanism that mutual inhibition limits the number of stable bump states that can be simultaneously maintained. The second of them is a temporal mechanism based on the theta-gamma coupling of neural oscillation, in which a limited number of gamma cycles, representing distinct memorized items, can be fitted in each theta cycle~\citep{lisman2013theta}. Interestingly, although our model does not explicitly represent temporal oscillations, it is compatible with the interpretation that the observed correlations may reflect the synchrony or antisynchrony between different bump states. Thus, by taking into account correlated neural variability, our model hints toward a unifying perspective for the limit of working memory capacity based on spatiotemporal working memory representation.

These results also suggest an efficient computational strategy for memory storage by fluctuating neural activity, in which limited resources (i.e., neural spikes) are distributed across both space and time. Spatially, negative correlations increase memory capacity by reducing the distance between adjacent bumps (see Figure~\ref{fig:2_bump_compre}). Temporally, negative correlations can be interpreted as a form of antisynchrony that corresponds to temporally less concentrated bump activity states for each stored item. Conceptually, this is similar to time-division multiplexing recently proposed in a spiking neural network model~\citep{akam2014oscillatory,caruso2018single} and can also be understood as a form of sparse coding.

\begin{figure}[h]
    \centerline{\includegraphics[width=0.65\textwidth]{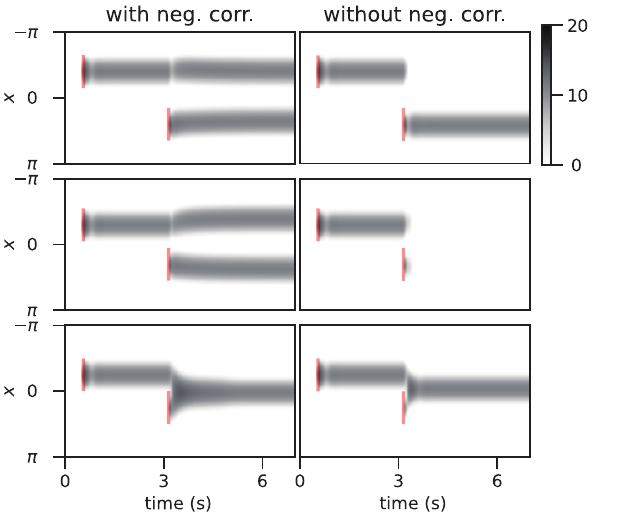}}
    \caption{\textbf{Negative correlations prevent forgetting of sequentially memorized items.}
    The left column shows the cases in which we keep the negative correlations. When the initial distance between two bumps is $0.8\pi$ (top panel) or $0.6\pi$ (middle panel), both of them remain persistent after the removal of the transient stimuli and reach a stable distance. If the initial distance between two bumps is sufficiently small ($0.5\pi$),the two bumps will merge into one bump (bottom panel). The right column shows the cases in which we clamp the negative correlations to zero. If the initial distance between two bumps is $d=0.8\pi$, the second bump is maintained, while the previously stored bump is quenched (top panel). If the initial distance between two bumps is $d=0.6\pi$, the two bumps will annihilate each other (middle panel). If the initial distance between two bumps is $d=0.5\pi$, the two bumps will merge into one bump (bottom panel). The system is initialized with zero mean and covariance, and two transient stimuli of duration $T=0.1$ s (marked by the red rectangles) are exerted at two locations sequentially. The grayscale indicates mean firing rate (sp/s).
    }
    \label{fig: seq_memory}
\end{figure}

We also find that negative correlations can improve the robustness of working memory by preventing the forgetting of a previously stored bump when a new stimulus arrives. To demonstrate this, we sequentially inject transient stimuli with durations of $T=0.1$ s and various spatial separations and investigate the resulting dynamics of bump attractors. As shown in Figure~\ref{fig: seq_memory}, if the initial distance between two bumps is sufficiently large ($0.6\pi$), both remain persistent after the removal of the transient stimuli and reach a stable distance under either attractive (top panel) or repulsive (middle panel) interactions. If the initial distance between two bumps is sufficiently small ($0.5\pi$), the two bumps will merge into one bump (bottom panel).
Next, we clamp the negative correlations to zero and observe the effect on the dynamics of sequentially memorized bump states (see the right column in Figure~\ref{fig: seq_memory}).
If the initial distance between two bumps is sufficiently large ($d=0.8\pi$, top panel), the second bump is maintained, while the previously stored bump is quenched. If the initial distance between two bumps is sufficiently small ($d=0.6\pi$, middle panel), the two bumps will annihilate each other. If the initial distance between two bumps is too small ($d=0.5\pi$, bottom panel), the two bumps will merge into one bump, which is the same outcome as the case with negative correlations. These results suggest that the negative correlations help to prevent forgetting due to competition from similar memorized items

\section{Discussion}
In this work, we have revealed the emergence of synergistic neural population codes and their mechanistic origin from nonlinearly coupled, correlated neural fluctuations in a bump attractor network. Our model demonstrates that complex spatial correlation structures can emerge from a bump attractor network model, as shown in Figure~\ref{fig:simple_bump_demo}, suggesting the need for closer experimental investigations of the pairwise correlations of neural population activity in the brain. As shown in Figure~\ref{fig:coding}a, the model additionally shows that global neural circuit parameters (such as the external input mean) can drastically change the dynamical regimes of bump attractor states with distinct noise correlation patterns, which can be interpreted as corresponding to different cognitive states. Specifically, our model predicts that behavioral error should be larger when the noise covariance is parallel to the tangent space of the ring manifold than when the noise covariance is orthogonal to the tangent space of the ring manifold. Future experimental studies could investigate this relationship between the neural population codes and behavioral performance using simultaneous recordings of cortical neurons during cognitive tasks. Furthermore, we reveal correlation induced dynamical interactions between two bump states, which lead to a stable bound state of two bumps as characterized by a local minimum in the effective interaction energy between them as shown in Figure~\ref{fig:2_bump_compre}. As a result, the interaction between two bumps, which represent the memorized items, can be attractive or repulsive depending on their distance (\ie, difference in the encoded features). To test our theory, future experimental studies could systematically investigate how behavioral errors during working memory tasks depend on the difference between memorized items or between one memorized item and a distractor stimulus. 
We also show that the negative correlation between the neural activity of two bumps are originated from mutual inhibitions in the recurrent network, a property often overlooked in previous firing rate models of bump attractors. These results thus highlight the importance of incorporating correlated neural variability in modeling studies of cortical dynamics.

The validity of analyzing stochastic neuronal dynamics through approximations up to second-order moments is supported by both empirical evidence and theoretical analysis. Through examining multielectrode array recordings in the vertebrate retina~\citep{schneidman2006weak},it is found that the strong collective behavior exhibited by a neural population is well accounted for by a model that captures the observed pairwise correlations but assumes no higher-order interactions, even in the case where the pairwise correlations are weak. A theoretical study~\citep{dahmen2016correlated} investigates the effect of higher-order cumulants to the dynamics in a stochastic binary neural network model. It is found that in a weakly correlated state, the contributions from higher-order cumulants to the network dynamics are significantly smaller than pairwise interactions. This suggests that the first two moments are fairly accurate at capturing the stochastic dynamics of the system. In summary, it can be argued that our MNN offers a minimalistic yet effective representation of fluctuating neural activity and is suitable for the purpose of modeling working memory in this work.

The MNN model presented in this study has a couple of limitations. First, the current version of MNN lacks the depiction of temporal statistics of neural activity, which may encode computationally useful information such as phase and time-delayed synchrony. Future studies are required to develop moment mappings that account for temporal statistics such as auto- and cross-covariances of neural spike trains~\citep{dahmen2016correlated}. 
Second, in order to simulate the dynamics of the MNN, it is necessary to store the covariance matrix of the system, which entails a quadratic space complexity. Furthermore, multiplying the covariance matrix with the connection weights results in a cubic time complexity. As a consequence, simulating large-scale neural systems could require significant computational resources. However, this trade-off is worthwhile considering that high dimensional joint probability distribution of neural activity is computationally intractable.

In the context of ring attractor networks, the MNN based on the gaussian field approximation can be generalized in a number of ways. First, we can consider a two-dimensional spatial structure instead of just one to incorporate more complex two-dimensional patterns~\citep{folias2005breathers,qimnn2022}. A straightforward 2D extension of the model considered in this paper is included in \si, where the firing covariability manifests as more complex patterns compared to the one-dimensional case. Second, this study considers stationary bump activity with correlated variability by using a model derived from the current-based leaky integrate-and-fire neurons without adaptation. Previous theoretical studies of bump attractors using firing rate models suggest that neural adaptations can lead to the formation of complex dynamical patterns such as propagating and rotating waves~\citep{folias2005breathers,qi2015dynamic,Wavesbumpspatterns,qi3863569fractional,senk2020conditions} and that short-term synaptic facilitation and depression can have a significant impact on the dynamics of bump diffusion~\citep{seeholzer2019stability,qi3863569fractional}. It is thus of great theoretical interest to further extend our model to incorporate neural adaptation, such as spike frequency adaptation, or synaptic depression, and investigate the influence of correlated neural fluctuations on the dynamics of the resulting wave patterns as well as its functional implications. Third, neural fluctuations may also cause bump states to drift away along the attractor manifold over time, resulting in further degradation of working memory. It has been shown that uncorrelated noise imposes a bound on the coding capability of bump attractor~\citep{burak2012fundamental}. The model presented in this paper may be extended to analyze the influence of correlated noise on this diffusive effect and its impact on working memory performance.

Although we have primarily focused on bump attractor networks in this paper, the general approach to modeling correlated neural variability using the MNN can be easily extended to other types of models. For instance, \cite{murray2017stable} proposed a parsimonious linear model to interpret the coexistence of dynamical coding and mnemonic coding in working memory. The network structure of this model can be applied to our MNN to gain a deeper understanding of the specific role played by firing covaribility in dynamical coding and mnemonic coding in working memory. Another extension is to divide the neural population into excitatory and inhibitory ones, as consistent with Dale’s principle~\citep{dale1935pharmacology}, for analyzing the dynamics and computational roles of firing covaribility in balanced network~\citep{van1996chaos} and in neural oscillations~\citep{pina2018oscillations}. Moreover, the MNN considered in this paper is derived from the simple leaky integrate-and-fire neuron model. The general framework of MNN can further incorporate more complex spiking neuron models with slow and fast synaptic dynamics, which are important in generating complex dynamics such as bursting activity~\citep{krahe2004burst}. Finally, the modeling framework based on the MNN also opens up new opportunities for investigating the role of neural correlations in synaptic plasticity, learning, and attention~\citep{KUSMIERZ2017170}.

\section*{Appendix}
\subsection{The moment neural network} 
To capture the nonlinear coupling of correlated neural activity, we employ a model known as the moment neural network derived from the following leaky integrate-and-fire neuron model~\citep{feng2006dynamics,lu2010gaussian}
\begin{align}\label{eq:LIF}
    \dfrac{dV_i(t)}{dt} = -LV_i(t)+I_i(t),\quad I_i(t)= \sum_{j=1}^Nw_{ij}S_j(t),
\end{align}
where $V_i$ is the membrane potential of the $i$th neuron, $L$ is the leak conductance, $I_i(t)$ is the total synaptic current, $w_{ij}$ is the synaptic weight from neuron $j$ to neuron $i$, and $S_j(t)$ is the spike train from pre-synaptic neurons. 
When $V_i$ reaches the firing threshold $V_{\text{th}}$
the neuron will release a spike which is transmitted to other connected neurons, and then $V_i$ is reset to the resting potential $V_{\text{res}}$ and enters a refractory period $T_{\text{ref}}$.
The functions $\phi_\mu$ and $\phi_\sigma$ together map the mean $\bar{\mu}$ and variance $\bar{\sigma}^2$ of the input current $I_i(t)$ to that of the output spikes according to~\cite{feng2006dynamics,lu2010gaussian}
\begin{align}
    &\phi_{\mu}: (\bar{\mu},\bar{\sigma}^2)\mapsto\mu, \quad\mu =  \dfrac{1}{T_{\rm ref} + \tfrac{2}{L}\int_{I_{\rm lb}}^{I_{\rm ub}} g(x) dx},\label{eq:mu}\\
    &\phi_{\sigma}: (\bar{\mu},\bar{\sigma}^2) \mapsto \sigma^2,\quad\sigma^2 = \tfrac{8}{L^2}\mu^3\textstyle\int_{I_{\rm lb}}^{I_{\rm ub}} h(x) dx,\label{eq:sigma}
\end{align}
where $T_{\rm ref}$ is the refractory period with integration bounds $I_{\rm ub}(\bar{\mu},\bar{\sigma}) = \tfrac{V_{\rm th}L-\bar{\mu}}{\sqrt{L}\bar{\sigma}}$ and
$I_{\rm lb}(\bar{\mu},\bar{\sigma}) = \tfrac{V_{\rm res}L-\bar{\mu}}{\sqrt{L}\bar{\sigma}}$. The constant parameters $L$, $V_{\rm res}$, and $V_{\rm th}$ are identical to those in equation~\eqref{eq:LIF}. The pair of Dawson-like functions $g(x)$ and $h(x)$ appearing in equations~\eqref{eq:mu} and \eqref{eq:sigma} are $g(x)=e^{x^2}\int_{-\infty}^x e^{-u^2}du$ and $h(x)=e^{x^2}\int_{-\infty}^x e^{-u^2}[g(u)]^2du$. 
The mapping $\psi$, which we refer to as the linear response coefficient, is equal to $\psi(\bar{\mu},\bar{\sigma}^2)=\tfrac{\partial\mu}{\partial\bar{\mu}}$ and it is derived using a linear perturbation analysis around correlation coefficient $\bar{\rho}_{ij}=0$~\citep{10.1038/nature06028,lu2010gaussian}. This approximation is justified as pairwise correlations between neurons in the brain are typically weak. Throughout this paper, we set neuron parameters to be $V_{\rm th}=20$ mV, $V_{\rm res}=0$ mV, $T_{\rm ref}=5$ ms, and $\tau=1/L=20$ ms.  
An efficient numerical algorithm is used for implementing the moment mappings~\citep{qimnn2022}.

\subsection{Network parameters}
We construct a ring network by setting distance-dependent synaptic weights with short-range excitation and long-range inhibition as follows,
\begin{align}\label{eq:weight}
  w_{ij} = \tfrac{2\pi}{N}[ w_E\kappa(x_i-x_j;d_E)-w_E\kappa(x_i-x_j;d_I)],
\end{align}
where $\kappa(x;d) = \exp[\tfrac{1}{d^2}(\cos x-1)]$, $w_E$ and $w_I$ control the strengths of excitatory and inhibitory synapses, and $d_E$ and $d_I$ control their spatial range, respectively. 
Throughout this paper, we set the synaptic weight parameters to be $w_E=15$, $w_I=6$, $d_E=0.5$, $d_I=1$, and $N=400$. 
We fix $\sigma_{\mathrm{ext}}^2=0.01$ throughout this study and systematically vary $\mu_{\mathrm{ext}}$.
We use Euler's method for simulating equations~\eqref{eq:mu_dynamics} and \eqref{eq:C_dynamics} with a time step of $\Delta t = 0.5\tau$.

\subsection{Linear Fisher information analysis}
To quantify the amount of information content encoded by a bump attractor, we calculate the linear Fisher information $I_{\mathrm{linear}}$ as follows~\cite{kohn2016correlations}. 
Given a bump state with mean firing rate $\mu(x\vert s)$ and firing covariability $C(x,{x}'\vert s)$ centered at $s$, the linear Fisher information can be calculated 
\begin{align}\label{eq:fisher_info}
    I_{\mathrm{linear}}(\Delta t)=  \mathbf{u}(s)^TC(s)^{-1}\mathbf{u}(s)\Delta t,
\end{align}
where $u_i(s) = \frac{d \mu_i(s)}{d s}$ and $\Delta t$ is the spike count time window. Note that $I_{\mathrm{linear}}(\Delta t)$ becomes approximately independent of $s$ due to the translation invarianace when the population size is large. 
For calculating $I_{\mathrm{linear}}(\Delta t)$, we apply centered finite difference to approximate $\mathbf{u}(s)$ and use Moore-Penrose pseudo-inverse for calculating $C(s)^{-1}$.

\subsection*{Acknowledgments}
Supported by STI2030-Major Projects (No. 2021ZD0200204); supported by Shanghai Municipal Science and Technology Major Project (No. 2018SHZDZX01), ZJ Lab, and Shanghai Center for Brain Science and Brain-Inspired Technology; supported by the 111 Project (No. B18015).

\bibliographystyle{apalike}
\providecommand{\noopsort}[1]{}\providecommand{\singleletter}[1]{#1}%

%\includepdf[pages={{},{},1,{},2,{},3,{},4,{},5,{},6,{},7,{},8,{},9}]{Supplementary.pdf}


\section*{Supplementary material}

# Supporting Information

# Self-organization of nonlinearly coupled neural fluctuations into synergistic population codes

Hengyuan Ma[1,†], Yang Qi[1,2,†], Pulin Gong[4], Jie Zhang[1,2], Wenlian Lu[1,2], Jianfeng Feng[1,2,3,*]

1 Institute of Science and Technology for Brain-inspired Intelligence, Fudan University, Shanghai 200433, China
2 Key Laboratory of Computational Neuroscience and Brain-Inspired Intelligence (Fudan University), Ministry of Education, China
3 Department of Computer Science, University of Warwick, Coventry, CV4 7AL, UK
4 School of Physics, University of Sydney, Sydney, NSW 2006, Australia

† These authors contributed equally.
∗ jffeng@fudan.edu.cn

## Linear Fisher information analysis

**Scaling behaviors of Fisher information in the large population size limit**

We calculate the ratio between the information and the population size. As shown in Fig. S1, this ratio converges to non-vanishing value in the limit of large population size for both state II and state III in Fig. 2(a), suggesting that information does not saturate.

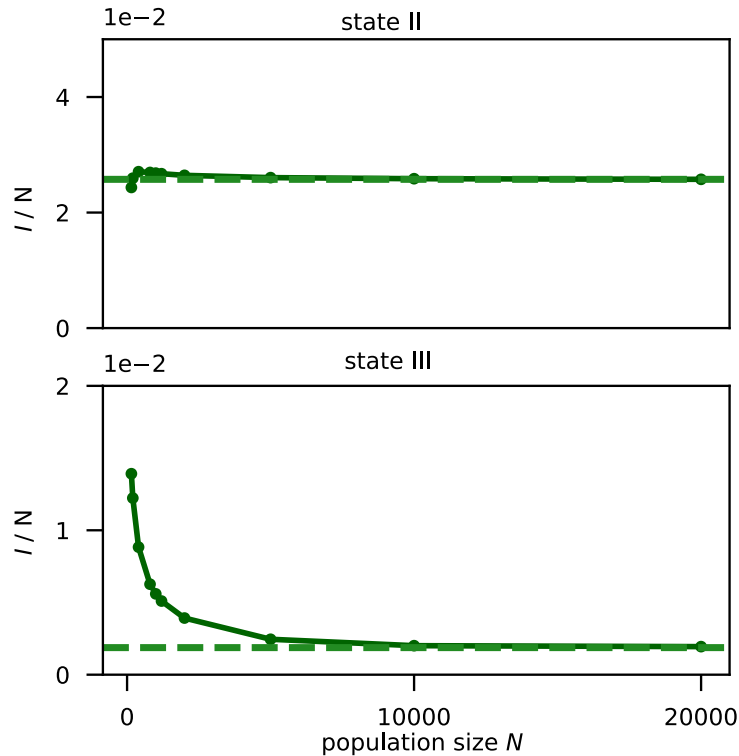

Figure S1: **Linear Fisher information does not saturate with the population size.** The ratio between the linear Fisher information rate $I(s)$ and the population size $N$ for state II (top panel) and state III (bottom panel). Both ratios converge to non-zero values, suggesting that the information does not saturate as the population size grows in both cases.

To gain further analytical insights to the scaling behavior of Fisher information in the limit of large population size, we considered simplified correlation structures based on state II and state III in Fig. 2(a) of our model. To proceed, we first

1

observe the symmetry in the differential covariance, $\mathbf{u}\mathbf{u}^T$, which has the following exact reflection symmetry

$$(\mathbf{u}_1)_i = -(\mathbf{u}_2)_{n+1-i}, \forall 1 \leq i \leq n, \tag{S1}$$

where we divide $\mathbf{u}$ into two blocks $\mathbf{u}_1, \mathbf{u}_2 \in \mathbb{R}^n$ as

$$\mathbf{u} = \begin{pmatrix} \mathbf{u}_1 \\ \mathbf{u}_2 \end{pmatrix}. \tag{S2}$$

Note that to simplify notations, we have assume that the population size $N = 2n$ is even and that bump center $s = 0$, though this does not affect the conclusion of the analysis in the limit of large population size. This symmetry suggests that we can analyze the interaction between noise covariance and differential covariance by applying above spatial division to the noise covariance

$$C = \begin{pmatrix} C_{11} & C_{12} \\ C_{21} & C_{22} \end{pmatrix}, \tag{S3}$$

where $C_{11}, C_{12}, C_{21}, C_{22} \in \mathbb{R}^{n \times n}$. We impose the following abstractions

$$(C_{11})_{ij} = (C_{12})_{i,n+1-j} = (C_{21})_{n+1-i,j} = (C_{22})_{n+1-i,n+1-j}, \tag{S4}$$

for state II, and

$$(C_{11})_{ij} = -(C_{12})_{i,n+1-j} = -(C_{21})_{n+1-i,j} = (C_{22})_{n+1-i,n+1-j}, \tag{S5}$$

for state III, for all $i \neq j$.

We first prove the non-saturation of the linear Fisher information for state II. Instead of directly calculating the linear Fisher information using Eq. (S31), we turn to construct a local linear decoder that estimate the bump location near $s$ and show that the variance $\sigma^2$ of this estimate diminishes as $N$ goes to infinity. We now construct a linear decoder as follows

$$\hat{s}(\mathbf{r}) = s + \frac{\mathbf{u}^T}{\mathbf{u}^T \mathbf{u}}(\mathbf{r} - \mu), \tag{S6}$$

where $\mu$ is the mean firing rate when the bump center is located at $s$, $\mathbf{r}$ is the firing rate of the system. The goal is to derive a lower bound on the Fisher information based on this local linear decoder. The variance of this estimation $\hat{s}(\mathbf{r})$ is

$$\text{var}[\hat{s}(\mathbf{r})] = \frac{\mathbf{u}^T}{\mathbf{u}^T \mathbf{u}} \text{cov}[\mathbf{r}] \frac{\mathbf{u}}{\mathbf{u}^T \mathbf{u}} = \frac{\Delta t}{\|\mathbf{u}\|^4} \mathbf{u}^T C \mathbf{u}. \tag{S7}$$

Using the division in Eq. (S3), we have

$$\mathbf{u}^T C \mathbf{u} = \mathbf{u}_1^T C_{11} \mathbf{u}_1 + \mathbf{u}_1^T C_{12} \mathbf{u}_2 + \mathbf{u}_2^T C_{21} \mathbf{u}_1 + \mathbf{u}_2^T C_{22} \mathbf{u}_2. \tag{S8}$$

Then, according to the symmetry condition in Eq. (S4), we obtain

$$\mathbf{u}^T C \mathbf{u} = \mathbf{u}_1^T C_{11} \mathbf{u}_1 - \mathbf{u}_1^T C_{11} \mathbf{u}_1 - \mathbf{u}_1^T C_{11} \mathbf{u}_1 + \mathbf{u}_1^T C_{11} \mathbf{u}_1 \tag{S9}$$

$$+ \sum_i (\mathbf{u}_1)_i^2 (C_{11})_{ii} + \sum_i (\mathbf{u}_1)_i (C_{12})_{ii} (\mathbf{u}_2)_i \tag{S10}$$

$$+ \sum_i (\mathbf{u}_2)_i (C_{21})_{ii} (\mathbf{u}_1)_i + \sum_i (\mathbf{u}_2)_i^2 (C_{22})_{ii} \tag{S11}$$

$$= 2 \sum_i (\mathbf{u}_1)_i^2 (C_{11})_{ii} + 2 \sum_i (\mathbf{u}_1)_i (C_{12})_{ii} (\mathbf{u}_2)_i \tag{S12}$$

$$\leq 2 \max_{i,j} |C_{ij}| \|\mathbf{u}_1\|^2 + 2 \max_{i,j} |C_{ij}| (\|\mathbf{u}_1\|^2 + \|\mathbf{u}_2\|^2) \tag{S13}$$

$$= 3 \max_{i,j} |C_{ij}| \|\mathbf{u}\|^2 \tag{S14}$$

Therefore, we have

$$\text{var}[\hat{s}(\mathbf{r})] \leq \frac{3 \max_{i,j} |C_{ij}| \Delta t}{\|\mathbf{u}\|^2}. \tag{S15}$$

For the synaptic weight defined by Eq. (10), the bump attractor converges to a finite spatial profile $\mu^*(x;s), C^*(x,x';s)$ in the limit of spatial continuum as $N \to \infty$, which implies that

$$\|\mathbf{u}\|^2 \sim \frac{N}{2\pi} \int_{-\pi}^{\pi} u^*(x;s)^2 dx = \frac{N}{2\pi} \|u^*\|_2^2, \tag{S16}$$

where $\|\cdot\|_2$ denotes the $L_2$-norm and $u^*(x;s) = \frac{\partial \mu^*(x;s)}{\partial x}$, and that $\max_{i,j} |C_{ij}|$ converges to the finite limit $C^*_{\max} = \max_{x,x'} |C^*(x,x';s)|$. Note that $\|u^*\|_2$ is a constant independent of $N$. Therefore, we have

$$\text{var}[\hat{s}(\mathbf{r})] \frac{6\pi C^*_{\max} \Delta t}{\|u^*\|_2^2 N}, \tag{S17}$$

which imposes a lower bound on the Fisher information according to the Cramér–Rao inequality,

$$I_{\text{linear}} \geq \frac{1}{\text{var}[\hat{s}(\mathbf{r})]} \frac{\|u^*\|_2^2}{6\pi C^*_{\max} \Delta t} N. \tag{S18}$$

Therefore, the linear Fisher information $I_{\text{linear}}$ increases at least linearly as the population size $N$ grows. In particular, the linear Fisher information does not saturate in state II.

In fact, the above conclusion holds for a wide class of the bump attractor states, *i.e.*, the information does not saturate, as long as the mean firing rate $\mu$ and the firing variability $C$ satisfy Eq. (S1) and Eq. (S4). The condition for the differential covariance [Eq. (S1)] is general, which is automatically satisfied if the bump is symmetrical. The condition for the noise covariance [Eq. (S4)] requires a mirror symmetry for the off-diagonal elements. As we have shown in our model, noise covariance structure qualitatively consistent with such mirror symmetry can naturally emerge from the recurrent dynamics of the bump attractor network. Intuitively, the mirror symmetry makes the $N^2 - 4N$ terms in the summation $\sum_{ij}(\mathbf{u})_i C_{ij}(\mathbf{u})_j$ cancel out following the reflection anti-symmetry condition of the differential covariance [Eq. (S11)-(S12)]. The remaining terms lead to a $1/N$ scaling in the variance $\text{var}[\hat{s}(\mathbf{r})]$ when $N$ is large enough.

For state III, we apply the same decoder [Eq. (S5)] as that used in state II and the spatial division [Eq. S3]. Then, according to the symmetry condition in Eq. (S5), the term $\mathbf{u}^T C \mathbf{u}$ in the estimate variance $\text{var}[\hat{s}(\mathbf{r})]$ in Eq. (S7) becomes

$$\mathbf{u}^T C \mathbf{u} = \mathbf{u}_1^T C_{11} \mathbf{u}_1 + \mathbf{u}_1^T C_{12} \mathbf{u}_2 + \mathbf{u}_2^T C_{21} \mathbf{u}_1 + \mathbf{u}_2^T C_{22} \mathbf{u}_2 \tag{S19}$$

$$= 4\mathbf{u}_1^T C_{11} \mathbf{u}_1 + 2\sum_i (\mathbf{u}_1)_i^2 (C_{11})_{ii} + 2\sum_i (\mathbf{u}_1)_i (C_{12})_{ii} (\mathbf{u}_2)_i. \tag{S20}$$

This does not clearly manifest whether or not the variance can diminish as $N \to +\infty$ due to the term $4\mathbf{u}_1^T C_{11} \mathbf{u}_1$.

Unfortunately, it is infeasible to determine a general statement about the scaling behavior of Fisher information under the condition in Eq. (S5). Nevertheless, we turn to prove the non-saturation of the linear Fisher information in a special case

$$(C_{11})_{ij} = (C_{22})_{ij} = \begin{cases} a, & \text{for } i = j \\ c, & \text{for } i \neq j \end{cases}, \tag{S21}$$

$$(C_{12})_{ij} = (C_{21})_{ij} = -c, \quad \text{for all } i, j, \tag{S22}$$

where $a > |c| \geq 0$ are constants. Note that we $c > 0$, this noise covariance matrix $C$ has the same sign as that in the differential covariance $\mathbf{u}\mathbf{u}^T$. The linear Fisher information for this case can then be calculated. First, we have

$$\begin{pmatrix} C_{11} & C_{12} \\ C_{21} & C_{22} \end{pmatrix}^{-1} = [(a-c)I_N + c \begin{pmatrix} O_n & -O_n \\ -O_n & O_n \end{pmatrix}]^{-1} \tag{S23}$$

$$= \frac{1}{a-c}[I_N - \frac{c}{(N-1)c + a} \begin{pmatrix} O_n & -O_n \\ -O_n & O_n \end{pmatrix}], \tag{S24}$$

where $O_n \in \mathbb{R}^{n \times n}$ is the matrix whose elements are all one. Thus, we have

$$I_{\text{linear}} = \mathbf{u}^T \begin{pmatrix} C_{11} & C_{12} \\ C_{21} & C_{22} \end{pmatrix}^{-1} \mathbf{u} \tag{S25}$$

$$= \frac{1}{a-c}[\|\mathbf{u}\|^2 - \frac{c}{(N-1)c + a}((\sum_i (\mathbf{u}_1)_i)^2 + (\sum_i (\mathbf{u}_2)_i)^2 - 2\sum_i (\mathbf{u}_1)_i \sum_i (\mathbf{u}_2)_i)] \tag{S26}$$

$$= \frac{1}{a-c}[\|\mathbf{u}\|^2 - \frac{c}{(N-1)c + a}(\sum_i (\mathbf{u}_1)_i - \sum_i (\mathbf{u}_2)_i)^2]. \tag{S27}$$

3

According to Eq. (S1), Eq. (S16) and by noting that

$$\left(\sum_i (\mathbf{u}_1)_i - \sum_i (\mathbf{u}_2)_i\right)^2 = \left(\sum_i |(\mathbf{u}_1)_i| + \sum_i |(\mathbf{u}_2)_i|\right)^2 \sim \frac{N^2}{4\pi^2} \|u^*\|_1^2, \tag{S28}$$

where $\|u^*\|_1 := \int_{-\pi}^{\pi} |u^*(x;s)| dx$ is the $L_1$ norm of $u^*(x;s)$, we have

$$I_{\text{linear}} \sim \frac{1}{a-c}\left[\frac{N}{2\pi}\|u^*\|_2^2 - \frac{c}{(N-1)c+a}\frac{N^2}{4\pi^2}\|u^*\|_1^2\right] \sim \frac{N}{4\pi^2(a-c)}[2\pi\|u^*\|_2^2 - \|u^*\|_1^2]. \tag{S29}$$

Finally, by invoking to Hölder's inequality, we have

$$\|u^*\|_1 = \|u^* \cdot 1\|_1 \leq \|u^*\|_2 \|1\|_2 = \sqrt{2\pi}\|u^*\|_2, \tag{S30}$$

and the equality holds if and only if $(u^*(x;s))^2$ is constant for $x \in [-\pi, \pi]$. Provided that $\mu(x;s)$ is even and twice differentiable, this leads to a spatially uniform solution of the mean firing rate, which is trivial. For general cases, the inequality Eq. (S30) is strict so that the linear Fisher information increases linearly as the population size $N$ grows.

In particular, if $c > 0$, this is an instance where the noise covariance $C$ has the same sign as that of the differential covariance $\mathbf{u}\mathbf{u}^T$ without the saturation of information. Thus, it is not surprising to see that the linear Fisher information in state III does not saturate as shown in Fig. S1, in which case the both covariances have the same sign.

**Contribution to Fisher information by individual neurons**

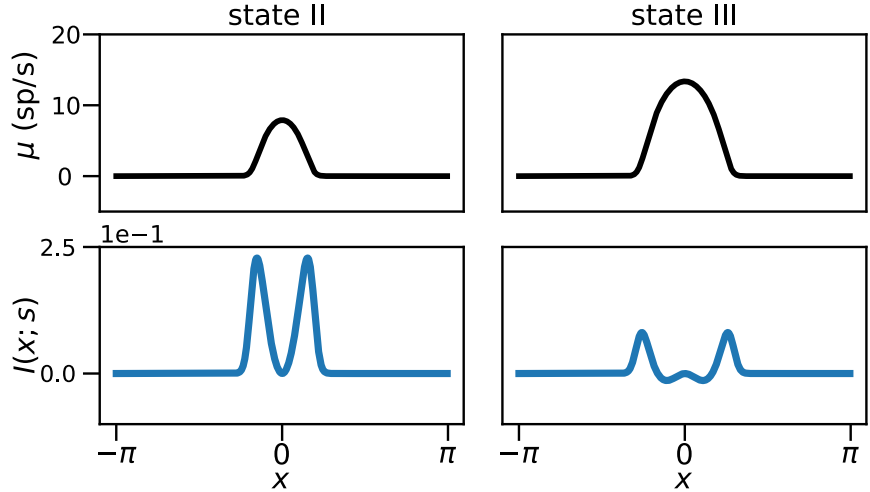

Figure S2: **The contribution to the linear Fisher information by individual neurons.** The mean firing rate (top panels) and the spatial distribution of the linear Fisher information rate $I(x;s)$ (bottom panels) for state II ($\mu_{\text{ext}} = 0.925$) and III ($\mu_{\text{ext}} = 0.932$). Here, the bump center $s = 0$.

Another property of interest regarding neural coding is the contribution to Fisher information by individual neurons in the network. We calculate the amount of linear Fisher information rate provided by each neuron located at $x_i$ as

$$I(x_i; s) = \sum_{j=1}^{N} u(s)_j (C(s)^{-1})_{ij} u(s)_i. \tag{S31}$$

As shown in Fig. S2, we find that in state II, the neurons located at the boundary of the bump is more important for coding than other neurons, as consistent with previous models (Seung and Sompolinsky, 1993). However, in state III, the neurons near the center of the bump provide negative contributions to the linear Fisher information, which is previously unpredicted, suggesting that some neurons may be harmful for coding.

## Bump interaction analysis

To investigate the mechanism of the interaction between two bumps near the stable bound state, we focus on the dynamics of the one of the bumps (free moving bump) and freeze that of the other bump (the clamped bump) which

acts on the free bump as if it is an external feedforward input. To facilitate the analysis, we set the origin to be the center between the two bumps at equilibrium distance. Due to the translation symmetry of the model, the origin chosen does not affect the generality of the proceeding analysis. Denote the mean firing rate and the firing covariability of the two-bump configuration in Fig. 4(b) as $\mu = (\mu_i)$ and $C = (C_{ij})$. We decompose the neural activity state based on the neurons' spatial positions $x_i$ such that

$$\mu = \mu^{\text{f}} + \mu^{\text{c}}, \tag{S32}$$

where $\mu_i^{\text{f}} = \mu_i H(-x_i)$ and $\mu_i^{\text{c}} = \mu_i H(x_i)$ with $H$ denoting the Heaviside step function. The Heaviside step function $H(x) = 1$, if $x \geq 0$, and zero otherwise. We also decompose $C$ accordingly as follows

$$C = C^{\text{ff}} + C^{\text{fc}} + C^{\text{cf}} + C^{\text{cc}}, \tag{S33}$$

where

$$C_{ij}^{\text{ff}} = C_{ij} H(-x_i) H(-x_j), \quad C_{ij}^{\text{fc}} = C_{ij} H(-x_i) H(x_j), \tag{S34}$$

$$C_{ij}^{\text{cf}} = C_{ij} H(x_i) H(-x_j), \quad C_{ij}^{\text{cc}} = C_{ij} H(x_i) H(x_j). \tag{S35}$$

This leads to the following modified equation for the mean firing rate of the free bump

$$\tau \frac{d\mu^{\text{f}}}{dt} = -\mu^{\text{f}} - \mu_{\text{effective}} + \phi_\mu(\bar{\mu}^{\text{f}} + \bar{\mu}_{\text{effective}}, \bar{C}^{\text{ff}} + \bar{C}_{\text{effective}}), \tag{S36}$$

where the synaptic current is

$$\bar{\mu}_i^{\text{f}} = \sum_k w_{ik} \mu_k^{\text{f}} + \mu_{\text{ext}}, \tag{S37}$$

the synaptic current covariability is

$$\bar{C}_{ij}^{\text{ff}} = \sum_{k,l} w_{ik} C_{kl}^{\text{ff}} w_{jl} + \sigma_{\text{ext}}^2 \delta_{ij}, \tag{S38}$$

where $\mu_{\text{effective}} = \mu^{\text{c}}$, $C_{\text{effective}} = C^{\text{fc}} + C^{\text{cf}} + C^{\text{cc}}$, $\bar{\mu}_{\text{effective}} = \sum_k w_{ik} \mu_k^{\text{c}}$, and $\bar{C}_{\text{effective}} = \sum_{k,l} w_{ik} (C_{kl}^{\text{fc}} + C_{kl}^{\text{cf}} + C_{kl}^{\text{cc}}) w_{jl}$. We refer to $C^{\text{fc}}$ and $C^{\text{cf}}$ as the inter-bump covariances, and $C^{\text{ff}}$ and $C^{\text{cc}}$ as the intra-bump covariance. In our model, we find that the inter-bump correlations are negative. We apply the following projection methods as inspired by (Burak and Fiete, 2012). Denote the activity of the free bump in the stable bound state as

$$\mu_i^{\text{f}} = \mu^{\text{f}}(x_i|s^{\text{f}}), \quad C_{ij}^{\text{ff}} = C^{\text{ff}}(x_i, x_j|s^{\text{f}}), \tag{S39}$$

where $s^{\text{f}}$ is the peak location of the free bump. Assume that its shape does not change significantly with small spatial deviations $z$ in the vicinity of the stable location $s^{\text{f}}$ such that

$$\mu_i^{\text{f}}(z) = \mu^{\text{f}}(x_i|s^{\text{f}} + z). \tag{S40}$$

The spatial deviation $z$ also shifts the intra-bump covariance of the free bump and the inter-bump covariances while leaving the intra-bump covariance of the clamped bump unchanged, that is

$$C_{ij}^{\text{ff}}(z) = C^{\text{ff}}(x_i, x_j|s^{\text{f}} + z), \tag{S41}$$

$$C_{ij}^{\text{fc}}(z) = C^{\text{fc}}(x_i, x_j|s^{\text{f}} + z, s^{\text{c}}), \tag{S42}$$

$$C_{ij}^{\text{cf}}(z) = C^{\text{cf}}(x_i, x_j|s^{\text{c}}, s^{\text{f}} + z), \tag{S43}$$

$$C_{ij}^{\text{cc}}(z) = C^{\text{cc}}(x_i, x_j|s^{\text{c}}). \tag{S44}$$

We next calculate the neutrally stable left eigenspace along the attractor manifold represented by the mean firing rate of the free moving bump as $v = a \frac{d\bar{\mu}^{\text{f}}}{dz}$, parameterized by the bump's spatial deviation $z$, where $a = 1/\langle \frac{d\bar{\mu}^{\text{f}}}{dz}, \frac{d\mu^{\text{f}}}{dz} \rangle$ is the normalization coefficient such that $\langle v, \frac{d\mu^{\text{f}}}{dz} \rangle = 1$ ($\frac{d\mu^{\text{f}}}{dz}$ is the right eigenvector of the system). Substitute above definition for $\mu_i^{\text{f}}$ into Eq. (1), we get

$$\tau \frac{\partial \mu_i}{\partial t} = -\tau u_i' \frac{dz}{dt}$$

on the left-hand side, and $\phi_\mu(\bar{\mu}_i, \bar{C}_i)$ on the right-hand side. Then we numerically project the system state onto the space spanned by $v$, to obtain the equation of motion for the spatial deviation $z$

$$\frac{dz}{dt} = \langle v, -\mu(z) + \phi(\bar{\mu}(z), \bar{C}(z)) \rangle := b(d), \tag{S45}$$

5

where the distance between the two bumps is $d = s^{\text{c}} - s^{\text{f}} - z$. The projection method is valid near the stable equilibrium at $z = 0$. We interpret $b(d)$ as the influence of the clamped bump on the motion of the free moving bump along the attractor manifold when their distance is $d$. Next, we define the energy landscape induced by this effective interaction as

$$E(d) = -\int_\pi^d b(x)dx. \tag{S46}$$

The effective interactive energy is calculated in two cases, one with all covariances included and the other with the inter-bump covariance $C^{\text{fc}}$ and $C^{\text{cf}}$ fixed at zero. All derivatives are computed numerically with centered finite difference.

## The origin of the negative correlations

To gain further theoretical understanding about the origin of negative correlations between two bumps, we consider the overall interactions between the neural populations of these two bumps, which can be roughly characterized by two neural masses

$$\mu = \begin{pmatrix} \mu_1 \\ \mu_2 \end{pmatrix}, C = \begin{pmatrix} c_{11} & c_{12} \\ c_{21} & c_{22} \end{pmatrix}, \tag{S47}$$

with self-excitatory and mutually inhibitory synaptic weights

$$W = \begin{pmatrix} w_{11} & w_{12} \\ w_{21} & w_{22} \end{pmatrix} \tag{S48}$$

satisfying

$$w_{11}, w_{22} > 0, \quad \text{(self-excitation)}, \tag{S49}$$
$$w_{12}, w_{21} < 0, \quad \text{(mutual inhibition)}. \tag{S50}$$

The quantities $(\mu_1, c_{11})$ and $(\mu_2, c_{22})$ represent the overall population activity of two bumps, and $c_{12} = c_{21}$ are the covariances between them. We can prove that under a mild condition of the weight matrix, two populations should be negatively correlated, *i.e.*, $c_{12} < 0$. Based on Eq. (2), the steady state covariance should satisfy

$$c_{12} = \psi(\bar\mu_1, \bar\sigma_1^2)\psi(\bar\mu_2, \bar\sigma_2^2)\bar c_{12}, \tag{S51}$$

where $\bar\sigma_1^2 = \bar c_{11}, \bar\sigma_2^2 = \bar c_{22}$, and

$$\bar c_{12} = (wcw^T)_{12} = w_{11}w_{21}c_{11} + w_{12}w_{21}c_{21} + w_{11}w_{22}c_{12} + w_{12}w_{22}c_{22}. \tag{S52}$$

Substitute Eq. (S52) to Eq. (S51) to obtain

$$[1 - \psi(\bar\mu_1, \bar\sigma_1^2)\psi(\bar\mu_2, \bar\sigma_2^2)(w_{11}w_{22} + w_{12}w_{21})]c_{12} = \psi(\bar\mu_1, \bar\sigma_1^2)\psi(\bar\mu_2, \bar\sigma_2^2)(w_{11}w_{21}c_{11} + w_{12}w_{22}c_{22}). \tag{S53}$$

Since $c_{11}, c_{22} > 0$, $w_{11}w_{22}, w_{12}w_{21} > 0$, $w_{11}w_{21}, w_{12}w_{22} < 0$, and

$$\psi(\bar\mu, \bar\sigma^2) = \frac{\partial \mu}{\partial \bar\mu} > 0, \quad \text{for all } \bar\mu, \bar\sigma > 0 \tag{S54}$$

we have $c_{12} < 0$ as long as

$$\psi(\bar\mu_1, \bar\sigma_1^2)\psi(\bar\mu_2, \bar\sigma_2^2)(w_{11}^2 + w_{12}^2) < 1, \tag{S55}$$

which is a mild condition satisfied by the synaptic weights and numerical solutions found in our model (Lu et al., 2010; Qi, 2022).

## Additional phase transitions for two-bump solutions

Apart from the patterns demonstrated in Fig. 4, other possible patterns of coupled bumps attractor state may also appear as $\mu_{\text{ext}}$ varies. As shown in Fig. S3(a), when $\mu_{\text{ext}}$ is strong enough coupled bump states emerge with positive intra-bump correlations and negative inter-bump correlations [state A in Fig. S3(a)]. Similar to the 1-bump case, negative correlations emerge between neurons on the opposite sides within each bump [state B in Fig. S3(a)], and the same pattern of correlations also appears between two bumps. Further increasing $\mu_{\text{ext}}$ causes the merging of two bumps, leading to a single bump with two peaks [state C in Fig. S3(a)]. Note that these merged bump states can co-exist with

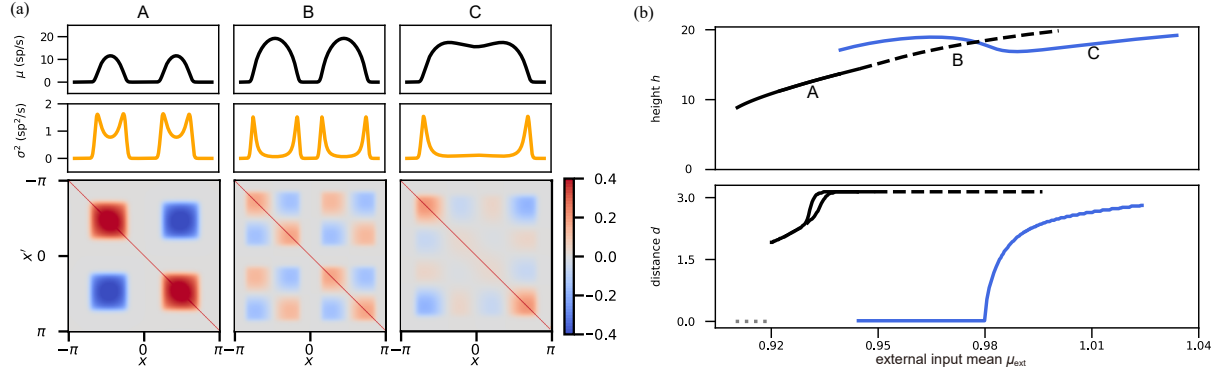

Figure S3: **Additional phase transitions for two-bump solutions.** (a) Mean firing rate $\mu(x)$, firing variability $\sigma^2(x)$ and the correlation coefficients $\rho(x, x')$ of neural activity states for varying external input mean $\mu_{\text{ext}}$, including when the inter-bump correlations are negative (state A, $\mu_{\text{ext}} = 0.940$), when the inter-bump correlations contain both positive an negative values (state B, $\mu_{\text{ext}} = 0.992$), and when the two bumps partially overlap each other (state C, $\mu_{\text{ext}} = 1.000$). (b) The height $h$ (top panel) and distance $d$ of two bumps (bottom panel) for varying external input mean $\mu_{\text{ext}}$, revealing distinct phase transitions between different activity states shown in (a).

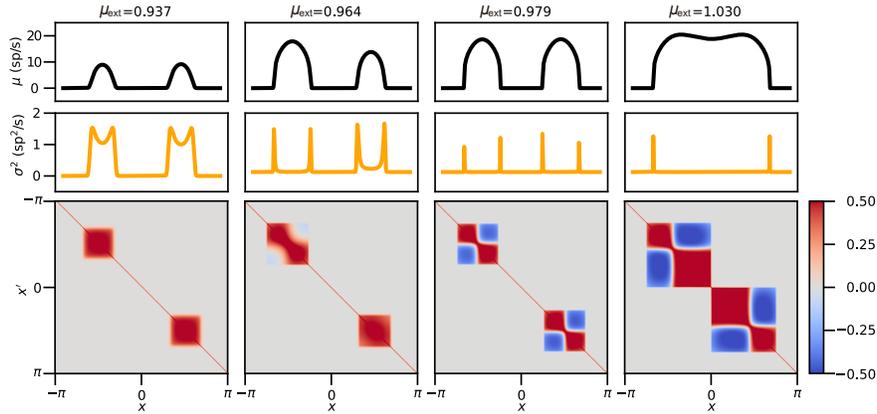

Figure S4: **The dynamics of two bumps when intra-bump correlation are fixed to zero.** The plots show the mean firing rate $\mu(x)$ (top panels), firing variability $\sigma^2(x)$ (middle panels) and the correlation coefficients $\rho(x, x')$ (bottom panels) for varying external input mean $\mu_{\text{ext}}$. In all cases, the distance between bumps are $d = \pi$, the maximal distance permitted by the model.

the state B when $\mu_{\text{ext}}$ takes moderately large values. In this case, when the initial positions of two bumps are close enough, the system evolves into the state C, and remains in state B otherwise.

Figure S3(b) shows the height $h$ of and the distance $d$ between two bumps under for varying $\mu_{\text{ext}}$. It is observed that both $h$ and $d$ increase as $\mu_{\text{ext}}$ increases for the two-bump case (black line), while the variation of $h$ is non-monotonic for the merged bump case (blue line). Note that when $\mu_{\text{ext}} \in [0.930, 0.938]$, the equilibrium distance between two bumps has two stable values, depending on their initial distance.

As shown in Fig. S4, we clamp the intra-bump correlations to zero, and find that the resulting distance between two bumps is always at $d = \pi$, the maximal distance permitted by the model, leading to reduced memory capacity.

## Mnemonic and dynamic coding during working memory

Murray et al. (2017) analyzed the temporal dynamics of the neural activity during visual working memory tasks and found that during the cue epoch (the presentation of the transient stimulus) a mnemonic coding space coexists with a dynamic coding space, and that during the delay epoch, the dynamic coding space decays while the mnemonic coding

7

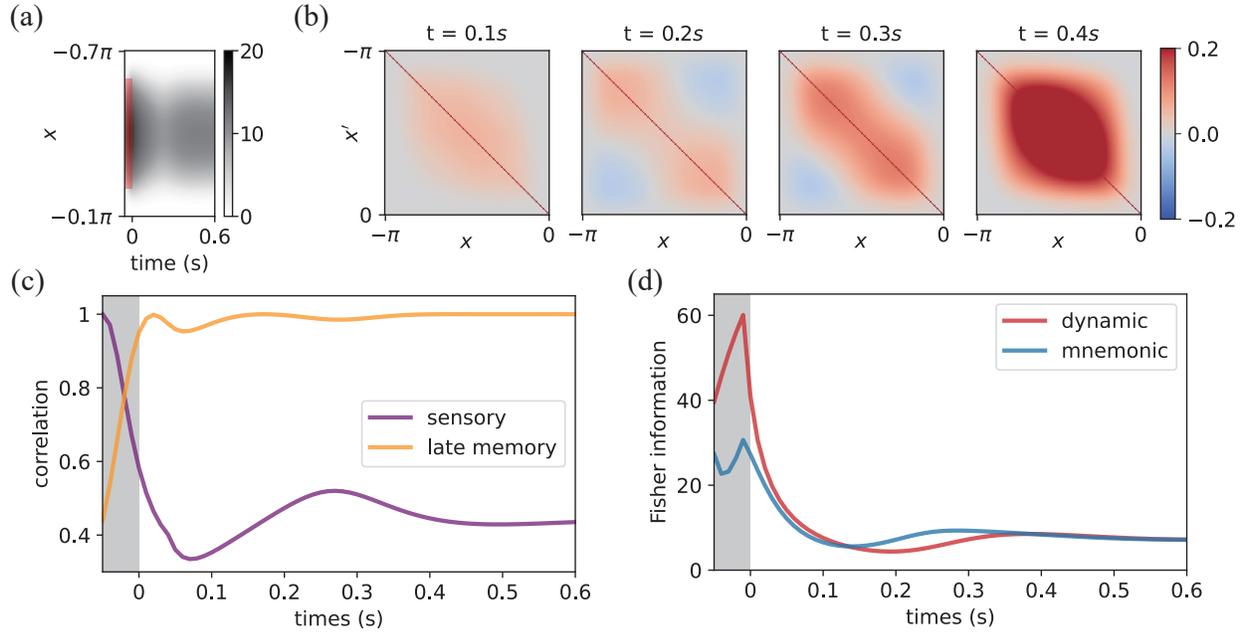

Figure S5: **Dynamic coding of the working memory at the state II in Fig. 2 for the transient stimulus** (a) The dynamics of mean firing rate $\mu(x)$ at the cue presentation (red rectangle) and delay epoch. (b) The dynamics of correlation coefficients $\rho(x, x')$ at the cue epoch and delay epoch. (c) Population correlation between states and two reference states: sensory state (purple) and late memory state (orange). The sensory state is defined by the start of the cue epoch and the late memory state by the end of the delay epoch. The gray area represents the cue epoch. (d) Fisher information over time for the mnemonic (blue) and dynamic (red) coding subspaces.

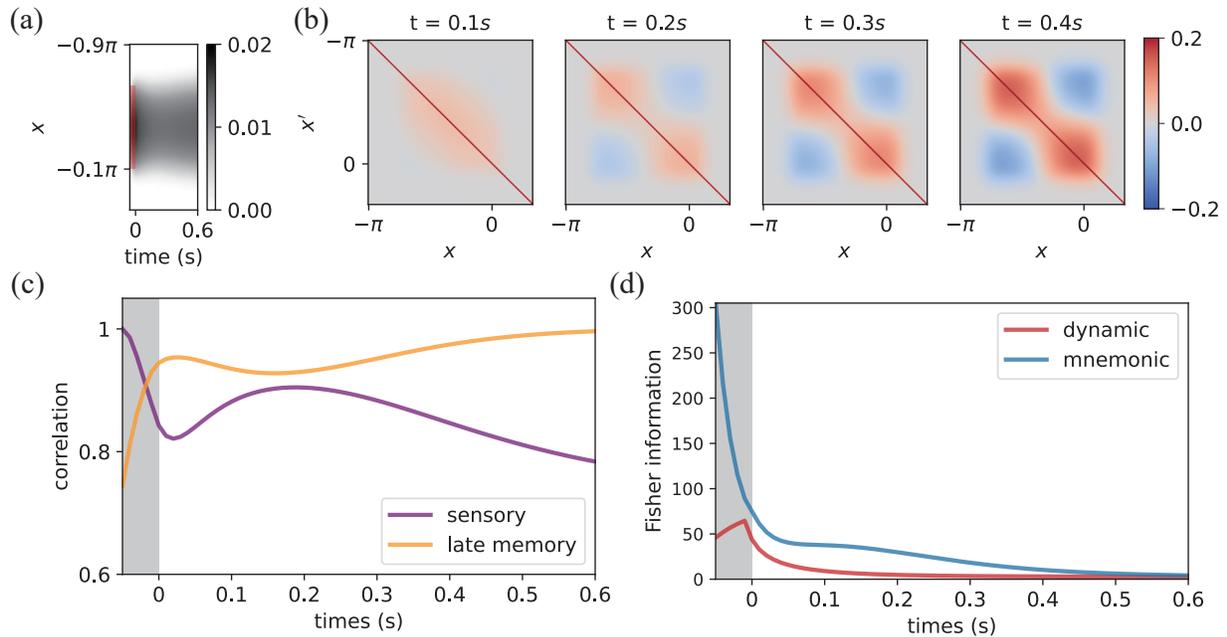

Figure S6: **Dynamic coding of the working memory at the state III in Fig. 2 for the transient stimulus** (a) The dynamics of mean firing rate $\mu(x)$ at the cue presentation (red rectangle) and delay epoch. (b) The dynamics of correlation coefficients $\rho(x, x')$ at the cue epoch and delay epoch. (c) Population correlation between states and two reference states: sensory state (purple) and late memory state (orange). The sensory state is defined by the start of the cue epoch and the late memory state by the end of the delay epoch. The gray area represents the cue epoch. (d) Fisher information over time for the mnemonic (blue) and dynamic (red) coding subspaces.

8

persists. To test whether our model is consistent with these results, we investigate the dynamics of mean firing rate and firing covariability of the MNN during the cue and delay epochs following Murray et al. (2017). We record the bump dynamics during and after the presentation of a transient stimulus, as shown in Fig. S5(a)-(b). Note that the stimulus is presented in the same fashion as the first stimulus in Fig. 6.

To analyze the dynamics, we first calculate the population correlation between the bump activity states at different timepoints and those at two reference timepoints: a "sensory" state at the cue epoch and a "late memory" state at the end of the delay epoch. The time evolution of the population correlation is shown in Fig. S5(c). Given the mean firing rate of two states $\mu^a = (\mu_i^a)_{i=1}^N$ and $\mu^b = (\mu_i^b)_{i=1}^N$, their population correlations are calculated as (following Eq. S1 in the paper by Murray et al. (2017))

$$\rho = \frac{\text{cov}[\mu^a, \mu^b]}{\sqrt{\text{var}[\mu^a]\text{var}[\mu^b]}}, \tag{S56}$$

where cov and var are taken over neurons in the population. In other words, the population correlation measures the similarity between the mean firing rate at two different timepoints. We find that the correlation with the sensory state peaks at the cue epoch and then decays during the delay epoch, while the correlation with the late memory state increases across time and peaks at the delay epoch, as largely consistent with the finding by Murray et al. (2017) (see Fig. 1C in their paper). One difference is that the correlation with the late memory state peaks at the start of the delay period in the MNN, whereas in the experimental data it keeps growing before peaking at the end of the delay period. The cause of this discrepancy is analyzed as follows. In this paper, we focus on the internal properties of a stationary bump state without considering its diffusion along the ring manifold due to fluctuating neural activity. This diffusion would decrease the population correlation at two timepoints as their separation increases. If we take into account the diffusion, the population correlation between a given timepoint and the late memory state should then continue to rise and reach its highest point at the end of the delay epoch, in agreement with the experimental observations.

Murray et al. (2017) have also shown that the decoding accuracy based on the dynamic coding is significantly higher than that based on the mnemonic coding, and the accuracy based on both types of coding becomes similar during the delay epoch. The MNN provides a convenient way to analyzing the mnemonic and dynamic coding of neural population activity through calculating the linear Fisher information, which provides an upper bound to the accuracy obtainable by a linear decoder, without requiring explicit construction of such a decoder. At a timepoint $t$, the linear Fisher information of the dynamic coding is calculated as

$$I_t(s) = \mathbf{u}_t(s)^\text{T} C_t^{-1}(s) \mathbf{u}_t(s) \Delta t, \tag{S57}$$

where $\mathbf{u}_t(s) = \frac{d\mu_t(s)}{ds}$ represents the dynamic coding subspace and $\mu_t(s)$ is the mean firing rate at timepoint $t$. Similarly, the linear Fisher information of the mnemonic coding is calculated as

$$I_{t,T}(s) = \mathbf{u}_T(s)^\text{T} C_t^{-1}(s) \mathbf{u}_T(s) \Delta t, \tag{S58}$$

where $\mathbf{u}_T(s) = \frac{d\mu_T(s)}{ds}$ represents the mnemonic coding subspace and $\mu_T$ is the mean firing rate at the end of the delay epoch. In both cases, $C_t(s)$ is the firing covariability at timepoint $t$. As shown in Fig. S5(d), the linear Fisher information of the dynamic coding subspace is significantly higher than that of the mnemonic coding subspace during the cue epoch, and they converge to a similar value toward the end of the delay epoch. These results are qualitatively consistent with the analysis on decoding accuracy based on experimental data (Murray et al., 2017).

Interestingly, the opposite result is found when we repeat the same analyses on State III in which the stationary bump state exhibits both positive and negative correlations. As shown in Fig. S6, the Fisher information based on the mnemonic coding subspace now has a higher peak than the dynamic coding subspace. Above analyses based on the MNN thus indicate that the structure of neural correlation can have significant impact on the properties of dynamic coding during working memory. Moreover, the consistency in dynamic coding between experimental observation and State II of our model predicts that the brain might adopt the type of correlation structure in State II as a way to optimize coding efficiency during the cue period.

**Two dimensional case**

We generalize our model with correlated neural variability to two spatial dimensions. Consider $N = N_x \times N_y$ neurons on a square lattice with periodic boundary conditions, where neuron $i$ is assigned with two-dimensional spatial coordinates $(x_i, y_i)$ over $[-\pi, \pi]^2$. The synaptic weight between neuron $i$ and $j$ is

$$w_{ij} = \frac{4\pi^2}{N}[w_E \kappa(x_i - x_j, y_i - y_j, d_E) - w_I \kappa(x_i - x_j, y_i - y_j, d_I)], \tag{S59}$$

9

where $\kappa(x,y,d) = \exp[\frac{1}{d^2}(\frac{1}{2}(\cos x + \cos y) - 1)]$. We set $N_x = N_y = 40$, $w_E = 15$, $w_I = 4$, $d_E = 0.25$, $d_I = 0.5$. For numerical simulation, we use Euler's method with a time step of $0.5\tau$, and use the fast Fourier transform for spatial convolutions.

The two-dimensional neural circuit supports multiple localized bump states as shown in the left panel of Fig. S7(a). Similar to State II in Fig. 2 of the one-dimensional case, the spatial profile of the mean firing rate $\mu$ of each bump is unimodal, whereas that of the firing variability $C$ shows a dent in the center and is higher around the rim of the bump. We find that two-dimensional bumps exhibit more intricate spatial correlation structures as shown in the right panels of Fig. S7(a). Each plot represents the two-dimensional map of correlation coefficients between all neurons in the circuit and a selected neuron (marked with black dot) located at a distance of $0.1\pi$ from the center of the bump at an angle of $0°$, $45°$ or $180°$. For the example shown, two of the bumps exhibit positive intra-bump correlations and negative inter-bump correlations, analogous to State II in the one dimensional case, whereas one bump exhibits polarized correlations analogous to State III of the one dimensional case. Increasing $\mu_{\text{ext}}$ leads to almost the same pattern in terms of mean firing rate $\mu$ and firing varibility $\sigma^2$ but symmetry breaking in the correlation patterns, similar to the transition from State II to State III in [Fig. 2(a)]. Both the intra-bump and inter-bump correlations may contain positive and negative values, depending on the relative locations of the neurons.

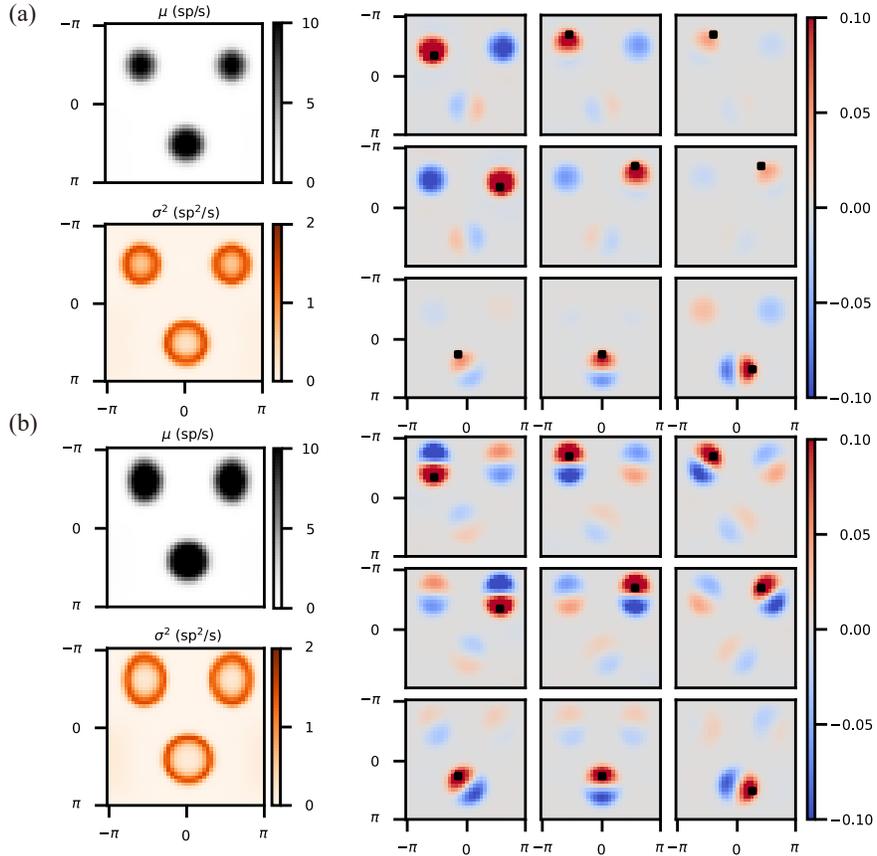

Figure S7: **Bump attractors with correlated variability in two spatial dimensions.** The displayed items are the two-dimensional maps of the mean firing rate $\mu$, firing varibility $\sigma^2$ and correlation coefficients between all neurons and neurons at locations marked by the black dots. The external input mean is set to be $\mu_{\text{ext}} = 0.945$ in (a), $\mu_{\text{ext}} = 0.950$ in (b).

**Detailed comparisons with stochastic rate-based neural network model**

Wimmer et al. (2014) analyzed single neuron recordings from the prefrontal cortex of the monkey brain and reported a series experimental observations during visual working memory tasks. They then used a rate-based bump attractor model driven by an uncorrelated additive white noise for theoretical interpretation on these experimental observations. In the following, we present a comparison between their model and the MNN model presented in this paper and highlight how the MNN offers alternative explanations to some of these experimental observations.

One of those experimental observations reported by Wimmer et al. (2014) is significant positive correlation between neuronal activity and behavioral response deviation during the delay period of working memory (see Fig. 4 in their paper). This rate-behavior correlation can be calculated in the MNN as follows. Let $\mathbf{n}$ be random spike count with mean $\mu \Delta t$ and covariance $C \Delta t$. Consider the decoder

$$\hat{\theta} = \text{Arg} \sum_{j=1}^{N} n_j \exp(i\theta_j), \tag{S60}$$

where $\theta_j \in [-\pi, \pi)$ is $j$-th neuron's preferred direction. The behavioral deviation following the definition by Wimmer et al. (2014) can then be modeled as

$$D_j = \alpha |\hat{\theta} - \theta^*|,$$

where $\theta^*$ denotes cue location and $\alpha = \text{sgn}(|\theta^* - \theta_j| - |\hat{\theta} - \theta_j|)$ indicates whether the behavioral response $\hat{\theta}$ is closer to, or further away from, the neuron's preferred location $\theta_j$. Without loss of generality, we fix the cue location to be $\theta^* = 0$. Note the absolute value calculates distance on a circle. The rate-behavior correlation is then defined as $c_j = \text{Corr}[D_j, n_j]$ (Wimmer et al., 2014), which is calculated by sampling $\mathbf{n}$ from the Gaussian distribution $\mathcal{N}(\mu \Delta t, C \Delta t)$. As shown in Fig. S8 (a), the bump activity in the MNN exhibits positive rate-behavior correlation as consistent with the analysis on experimental data (Fig. 4c in that paper). Moreover, distinct patterns are found for State II and State III in our model. For State II, the rate-behavior correlation shows a single peak centered at a neuron's preferred cue location. In contrast, the rate-behavior correlation for State III shows two peaks next to a neuron's preferred cue location, where the correlation is zero. The latter is consistent with experimental findings by Wimmer et al. (2014) (see Fig. 4e in their paper). Note that in the stochastic rate-based model considered by Wimmer et al. (2014) the rate-behavior correlation is primarily caused by the diffusion of bump attractor whereas in our case the correlation is due to the intrinsic fluctuation inside the bump (though the two mechanisms are not mutually exclusive).

Wimmer et al. (2014) have also shown that the spike count Fano factor (FF) of neuronal response shows systematic variation across different cue locations and that the FF for accurate and inaccurate trials differs significantly at the end of the delay period (see Fig. 6d in their paper). In the rate-based model, it is assumed that each neuron fires independent Poisson spikes with its instantaneous rate specified by the model. Under this assumption, the FF should always be equal to one unless the bump moves randomly causing the instantaneous firing rate to change over time. In the MNN, on the other hand, variable FF is produced by the intrinsic recurrent dynamics of the model, which is present even when the bump remains stationary. As shown in Fig. S8 (b), the FF for both State II and State III in the MNN exhibits a U-shaped tuning profile with values ranging from zero and one. This stimulus-dependent variability reduction is qualitatively consistent to that found in cortical neurons (Ponce-Alvarez et al., 2013). Note that the range of FF shown in Fig. S8 (b) is smaller than in experiments (Wimmer et al., 2014). This discrepancy could potentially be resolved if we consider the combined effect of the intrinsic variability of the bump state and its diffusive movement.

Wimmer et al. (2014) have further found that the noise correlation between pairs of neurons is negative for those pairs with dissimilar tuning when responding to a middle flank stimulus. This finding can be captured by State III of our MNN (left of Fig. 2c). For more convenient comparisons, we plot this result in the same format used by Wimmer et al. (2014) [Fig. S8 (c)]. No negative correlation is found in State II. Note that the noise correlation in MNN is quantitatively closer to experimental data (Fig. 7f in their paper) than in the rate-based model used by Wimmer et al. (2014) (Fig. S6g of their paper).

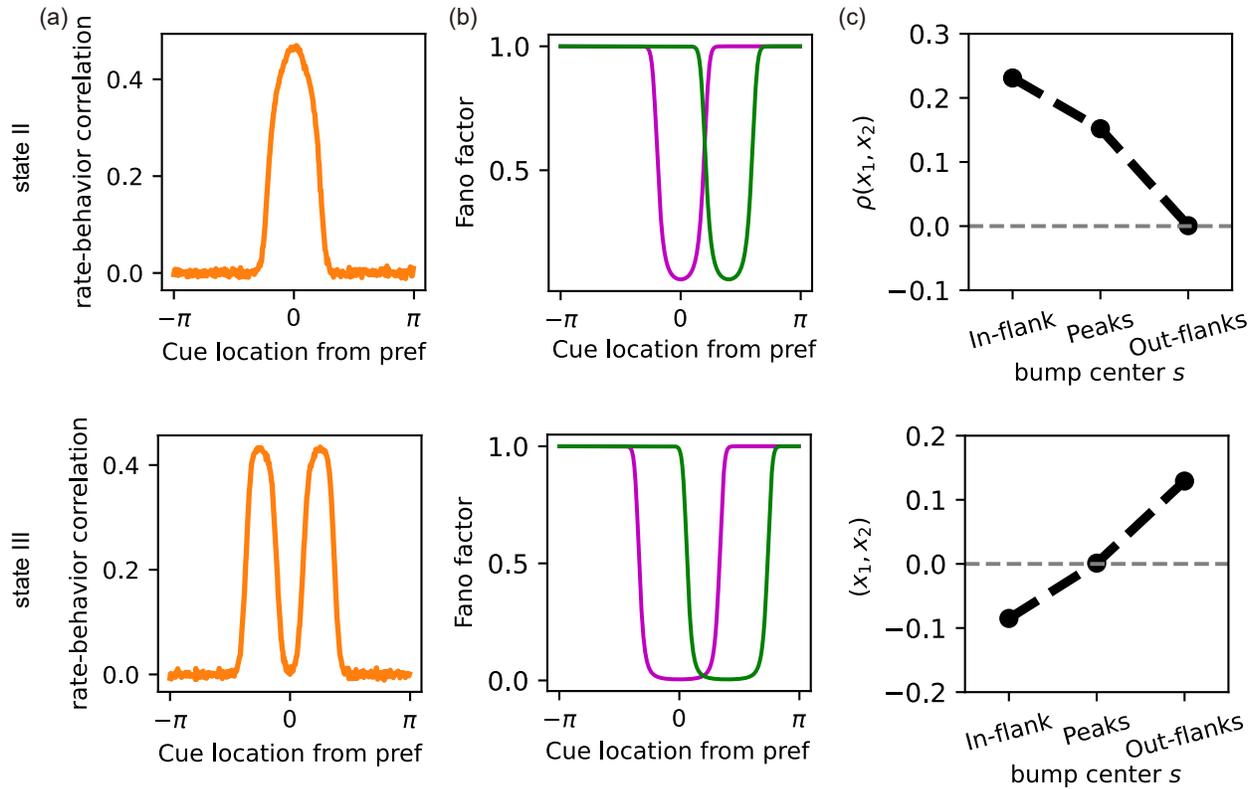

Figure S8: **Rate-behavior correlation, spike count Fano factor and neural pairwise correlation in the MNN model.** (a) Rate-behavior correlation $c_j$ as a function of cue location. Top row: State II; bottom row: State III. (b) Stimulus-dependent reduction of spike count Fano factor. Magenta lines represent the case with behavior response precisely matches the cue location and green lines trials with behavior response shifted from the cue location by $0.1\pi$. (c) Noise correlation of neuronal pair with dissimilar preferred cue locations. We select the bump center at $0.45\pi$ for representing the in-flank case, $0.5\pi$ for representing the peak case, and $0.56\pi$ for representing the out-flank case.



\end{document}